\documentstyle[aps,epsfig,twocolumn]{revtex}

\begin{document}

\title{Chaos in a closed Friedman-Robertson-Walker universe: an imaginary
approach}

\author{S.E. Jor\'as
\thanks{Email address: {\tt joras@joinville.udesc.br}}}
\address{\it Physics Department, State University of Santa Catarina, P.O. Box
631, Joinville, SC 89223-100, Brazil}
\author{T.J. Stuchi
\thanks{Email address: {\tt tstuchi@if.ufrj.br}}}
\address{\it Physics Institute, Federal University of Rio de Janeiro, P.O. Box 
68528, Rio de Janeiro, RJ 21941-972, Brazil}

\maketitle

\begin{abstract}
	In this work we study the existence of  mechanisms of the transition
to global chaos in a closed Friedmann-Robertson-Walker universe with
a massive conformally coupled scalar field.  We propose a
complexification of the radius of the universe so that the global
dynamics can be  understood. We show numerically the existence of
heteroclinic connections of the unstable and stable manifolds to
periodic orbits associated with the saddle-center equilibrium points.
We find two bifurcations which are crucial in creating non-collapsing
universes both in the real version and in the imaginary extension of the models.
 The techniques presented here can be employed in any cosmological model.

\end{abstract}

\section{Introduction}

	The characterization of chaos in gravitation and cosmology is
subtle. Positive Lyapunov exponents (which indicate an exponential
divergence of initially close trajectories in the phase space) must
not be used, since they can be made negative by a mere change of
coordinates \cite{george}. It is mandatory to use topologically
invariant procedures, such as the fractality of the attraction basin
boundaries, the analysis of Poincar\'e sections, or the structure of
periodic orbits (their bifurcations as well as homoclinic or heteroclinic
crossings of their invariant manifolds).

	On the other hand, it has been known that imaginary
 solutions may be necessary to the understanding of the dynamics of
 many chaotic systems \cite{biswas}.  As we shall see below, the somewhat
artificial complexification of the phase space, or actually, of the
scale factor  and its conjugated momentum, shines some light on the
chaotic nature of  the system in question. Although this extension can be
considered unphysical, it allows us to find a bifurcation
in the real Hamiltonian (\ref{ham}) with important consequences for the model
such as noncollapsing trajectories (universes) surrounded by chaotic regions
which lead to collapse after the trajectories spend a long period of time there.
Chaos is, thus, physically meaningful since it can be observed before the
universe collapses. That procedure is mandatory to the full understanding of
the onset of chaos in  real phase space.

\section{The model}

	Chaos in a Friedmann-Robertson-Walker (FRW) cosmological model filled with a
comformally coupled massive field has been considered  by many authors
 \cite{cornish,letelier,calzetta,blanco,bombelli,castagnino}.
 The metric is given by the line element
\begin{equation}
ds^2= a^2(\eta) \left\{ - d\eta^2 + \frac{1}{1-kr^2} dr^2 +
	r^2 \; d\Omega^2\right\}
\label{line}
\end{equation}

	In conformal time, the Friedmann equations can be cast
in the general form
\begin{equation}
2H = (p_\phi^2  + \phi^2 ) -  (p^2_a + a^2 ) +
m^2 a^2 \phi^2  = 0
\label{ham}
\end{equation}
where $a(\eta)$ is the scale factor.

	The above Hamiltonian would describe two coupled harmonic
oscillators if the sign of the scale-factor term (second parentheses)
 were not negative.  If one pushes the analogy, she/he would be forced
to admit  that one oscillator looses energy by {\it increasing} its
amplitude.

	Cornish and Levin \cite {cornish} took advantage of
the dissipative character in the matter sector of the system to show
chaos through the fractality  of the boundaries between basins  of collapse and
inflation. Their models consider also a cosmological constant as well as
a third coupled field. The same system was previously considered by
Calzetta \cite{calzetta}  who also exhibited evidence of chaos.
Both results were confirmed by Bombelli {\it et al.} \cite{bombelli}, with
the aid of the perturbative analitical method of Melnikov. Later on, Castagnino
{\it et al.} \cite{castagnino}, working in cosmological time, disregarded the
extension to negative values of the scale factor and thus
analytically found that there would be neither chaos nor inflation.

 In this paper we add a radiation component, whose energy density is given by
$\rho_r$, to the energy-momentum tensor. In this way we are able to relax the
constraint on the vanishing-energy condition \cite{stuchi}. From now on, this
constant will be absorbed by the Hamiltonian, i.e., $\tilde{H} \equiv 2H -
\rho_r$, and the tilde (~\~~) will be dropped to avoid cluttering of the
equations.

	In this paper we also acomplish a refined and nonperturbative analysis of the
numericalevolution of the two Hamiltonian systems in question in the complete
(real and imaginary) phase space.  We will be focusing our attention in the
existence of trajectories which do not collapse due the mechanism
provided by the (un)stable manifolds to periodic orbits in the $a>0$
range. Also the role played by two bifurcations are fully discussed
regarding the creation of KAM tori which {\bf do not cross the $(a=0)$ plane}.
One of them is shared by both the imaginary and the original (real) FRW
models. So, we have found noncollapsing universes which were not shown
in the previous discussions of this model. Moreover, there are
regions of regular and chaotic behavior which never collapse.
These regions are surrounded by chains of islands of various order
which allow some sticking of the universes, that is, they
wander for a long time before collapsing. It would be
interesting to do some further work to compaire with the
findings of Castagnino {\it et al.} \cite{castagnino} working
in cosmological time.

	The  complexification of the model  is done in Sec.~\ref{complex}.  The
fixed points and dynamical analysis  are discussed in Sec.~\ref{fixed}.
Section~\ref{qualit} brings forward the qualitative features of the extended
phase space, namely how caos sets in. In  Sec.~\ref{num} we present the
numerical  results and, in Section \ref{real}, we discuss the features of the
real (as opposed to complex) problem.

\section{Analytical extension}
\label{complex}

	Having in mind the role played by complex solutions of
the equation of motion in chaotic mechanics \cite{biswas}, we
propose the phase space must be complexified. That is, we will make
\begin{equation}
a(\eta) \rightarrow i \, a(\eta) \quad {\rm and} \quad p_a \rightarrow \,-i
\,p_a
\end{equation}
from now on. Indeed, as we shall see shortly, the system's
fixed-point structure shows all of its wealth, presenting a non trivial
dynamics.  This procedure is also used in Quantum Cosmology, where
 one deals with the tunneling of the wave function of the universe
(see, for instance, Ref.~\cite{quantum} and references therein).

	As far as the equations of motion are concerned, the above change
 takes  Hamiltonian (\ref{ham}) to the following form:
\begin{equation}
H=(p_\phi^2  + \phi^2) +  (p^2_a + a^2 ) - m^2 a^2 \phi^2
\label{ham2}
\end{equation}
which describes two coupled harmonic oscillators with an arbitrary
 total energy. Now one can rely on her/his physical intuition to
 describe the evolution of the system, since the two oscillators
 conserve energy and one can  only grow  at the other's expense.

	First, we analyse the structure of fixed points for an arbitrary
 value of the mass.   Then we present Poincar\'e sections of the
confined motion  in  spite of the fact that all such trajectories  go
to collapse. We  stress that the  dynamical  system defined by Eq.
(\ref{ham2}) is regular at $a=0$, contrary  to,   for instance,
Bianchi VIII and IX.  This tool, together with a numerical continuation of
periodic orbits and globalization of their stable-unstable manifolds, will be
largely used in our numerical exploration
of this model.  We are interested in understanding  the interplay of
the collapsing  trajectories  (either on tori or  irregular) with the
unstable and stable manifolds.  Then we will clearly see how escaping
orbits on $a>0$ avoid the collapse.

\subsection{Fixed points}\label{fixed}

	The fixed points of the ``rotated'' Hamiltonian (\ref{ham2}) are
easily found to be
\begin{eqnarray}
A &:& (0,0)\, ,\\
B_{\pm\pm} &:& (\pm 1/m, \pm 1/m)\nonumber
\end{eqnarray}
in $(a,\phi)$-coordinates. The origin is a stable equilibrium point, just as
before the rotation. The four nontrivial solutions $B_{\pm\pm}$, on the other
hand, have eigenvalues $\lambda_1^\pm=\pm \sqrt{3m^2-1}$ and
$\lambda_2^\pm=\pm i\sqrt{m^2+1}$, characterizing
four saddle-center fixed points.

 For energies higher than that of the fixed point, the motion can be
 split into two components: the center (in the direction of the
eigenvector  associated  to imaginary eigenvalues) and the saddle (to
real ones).  The former  leads  to stable periodic orbits around each
fixed point, if the energy is  restricted to this mode only.  On the
other hand,  if the energy is restricted to the other mode, the real
eigenvalues  will  lead the motion towards or away from the center
manifold  (see  Fig.~\ref{linsep}).


\subsection{Stable and unstable manifolds}
\label{qualit}

\begin{figure}[t]
\begin{center}
\epsfig{file=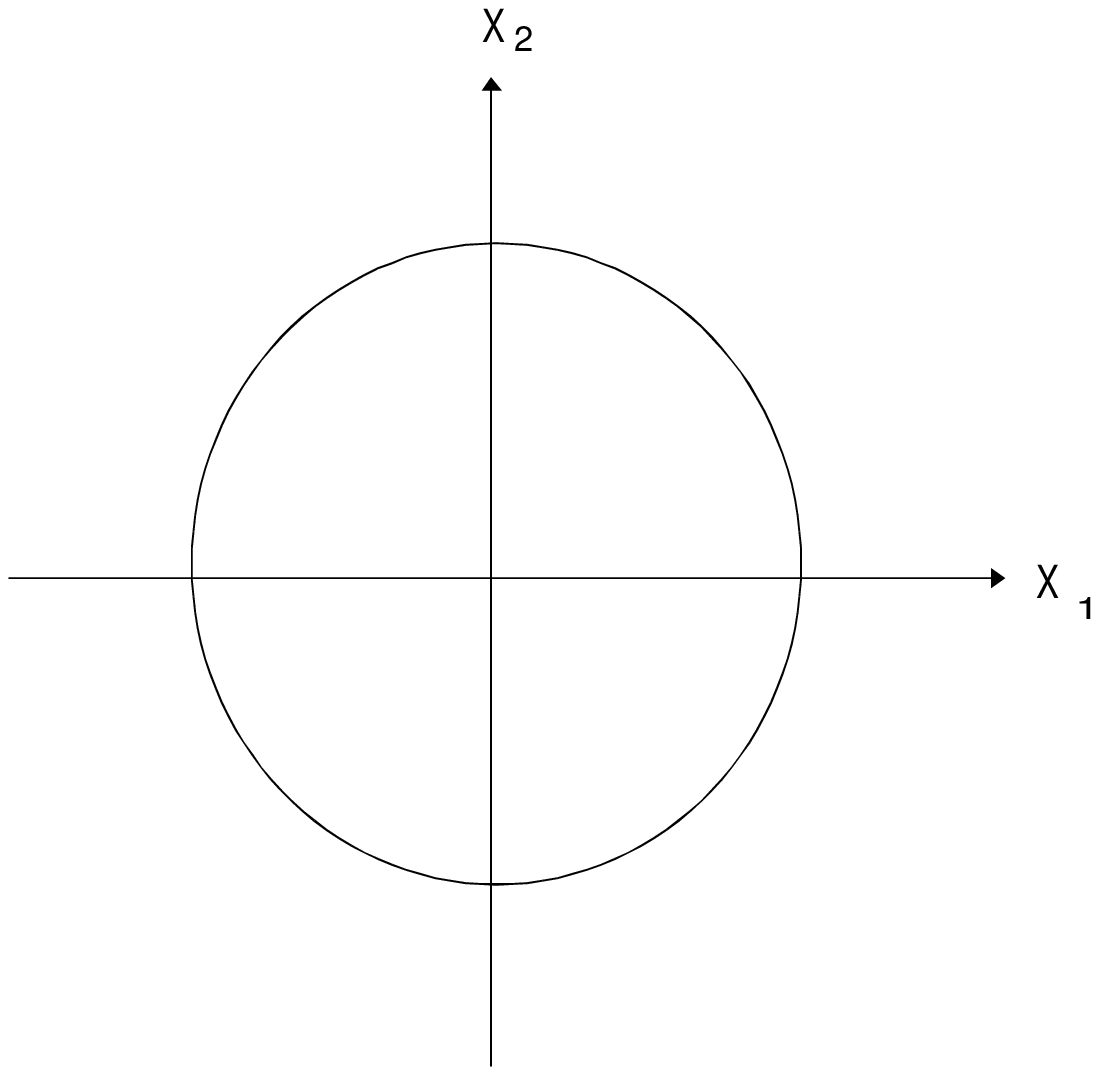,width=4cm,height=3cm}
\epsfig{file=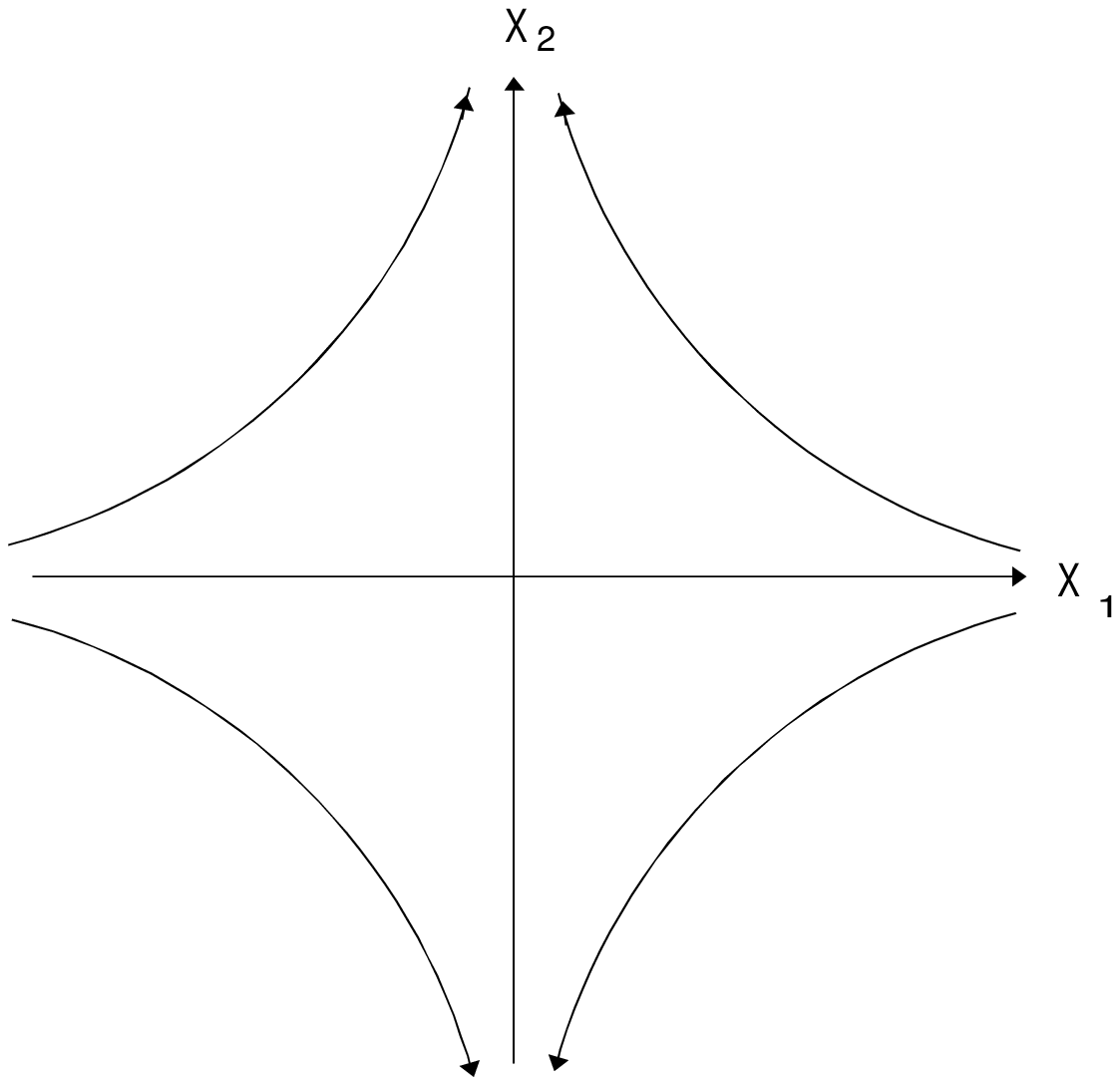,width=4cm,height=3cm}
\end{center}
\caption{Linear dynamics around $B_i.$ The left figure shows the
 center part while the right one, the saddle. }
\label{linsep}
\end{figure}

\begin{figure}[t]
\begin{center}
\epsfig{file=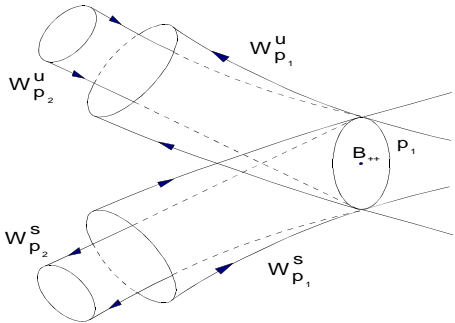,width=\columnwidth,height=4.5cm}
\end{center}
\caption{Resultant motion from the two directions sketched in the
previous figure. ${\rm W}^U_{P_1}$ stands for unstable manifold for
the periodic orbit $p_1$ around the fixed point $B_{++}$, while ${\rm
 W}^S_{P_1}$ stands for the stable one.}
\label{cylinder}
\end{figure}

	The combined motions discussed above define the stable and unstable
manifolds which look like cylinders emanating from each periodic
 orbit (see Fig.~\ref{cylinder}). The trajectories spirall along the
 cylinders towards and from the periodic orbit. Therefore, in the complete
linear dynamics the orbits sketched in Fig.~\ref{linsep} (top) are unstable.

	As we will see in Sec.~\ref{num}, the nonlinear analysis show that the
cylinders may intercept themselves (each other) at the so-called
homo(hetero)clinic conncetions.  Indeed,  the latter is exactly what happens in
the problem at hand.  Those  intersections  happen in regions among the
remnant KAM tori  from the  quasi-integrable  regime for low energy
values. Of course, they start from the energy  of  the saddle-center
fixed points occupying a tiny area of the  Poincar\'e section
  until  they eventually sweep away the whole tori  structure and
fill the whole phase space.  Therefore the chaotic  mechanism is
twofold: destruction of KAM tori due to resonances of  their
frequencies and the corrosion promoted from inside as the cylinders
get  larger and larger.

	The stable and unstable manifolds can be numerically calculated from
 the Poincar\'e sections of periodic orbits by a technique called
 numerical globalization.  We calculate the monodromy matrix --- the
variational equations evaluated after one period of the periodic
orbit, with the identity matrix as initial conditions.  Then we
determine the real eigenvector of the periodic orbits which yield the
linear directions  of the (un)stable manifolds.  Finally, the full
nonlinear cylinders can  be generated by taking a set of appropriate
initial conditions --- so that, in the Poincar\'e sections, we are
able to draw iterates which lie on closed curves corresponding to a section of
the trajectories forming the cylinders mentioned above \cite{stuchi2}.

	The symmetries of the problem allow us to save some computational
effort: from the unstable manifold of a periodic orbit around, say,
 $B_{++}$, we  can obtain all manifolds related to the periodic
orbits around the  other fixed points. There are eight such manifolds
altogether. However, only four of them intercept the chosen Poincar\'e section;
they are the unstable and stable  manifolds of periodic orbits
 symmetrical with respect to the origin.  The sections of the other four
manifolds can be obtained by  choosing the opposite sign of section crossing.

%

\subsection{Numerical results}
\label{num}

\begin{figure}[t]
\begin{center}
\vskip -.3cm
\epsfig{file=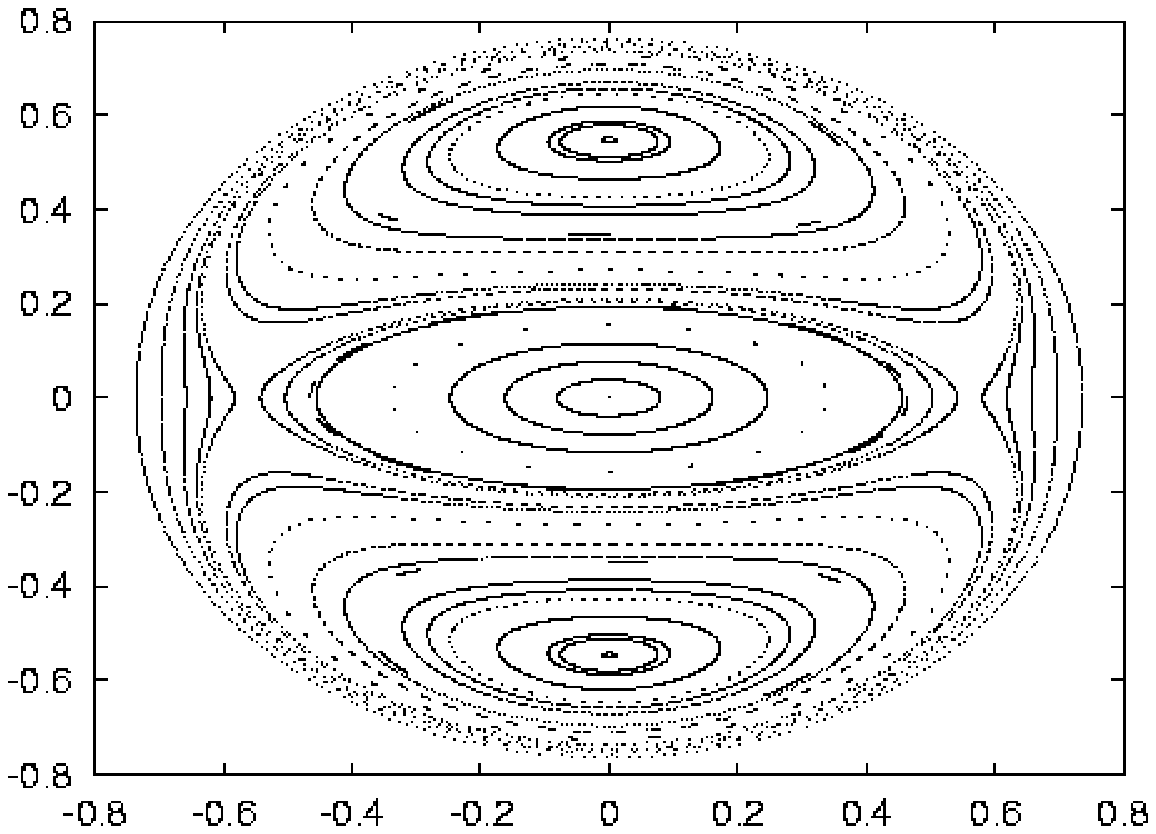,width=\columnwidth,height=5cm}
\epsfig{file=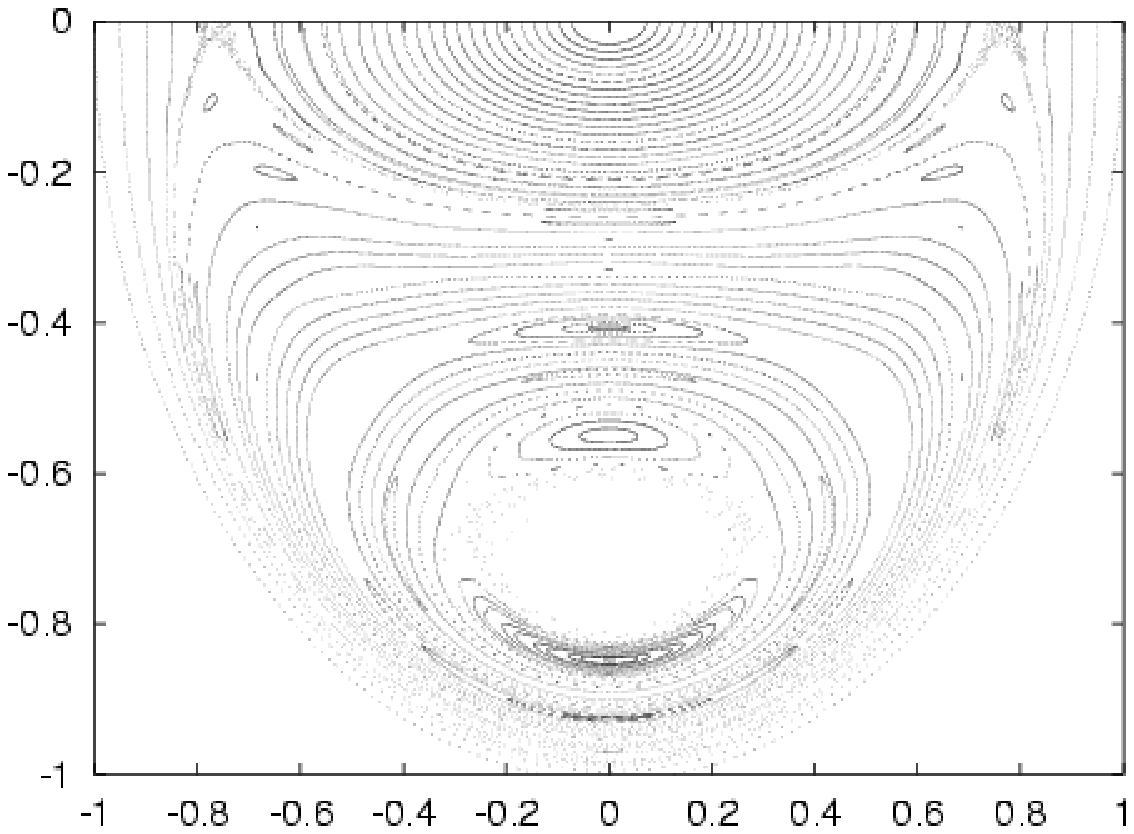,width=\columnwidth,height=5cm}
\end{center}
\caption{Poincar\'e section $\phi=0$, $\dot \phi>0$. (a) $E=0.3<E_B$; (b)
$E=0.511756758$, only $0.1$ larger than the energy threshold. The upper half is
a reflexion around the $p_a=0$ axis. Note the  gap in the KAM structure around
$a=0$ and $p_a=\pm 0.7$,  approximately.}
\label{SP}
\end{figure}

We have chosen the section $\phi=0$, $\dot \phi>0$. In
Fig.~\ref{SP}(a)  we can see this section for $E=0.3$, below the
energy threshold,  which is $E_B=0.5$ for $m=1.0$. The trajectories
are all in KAM tori with a very thin stochastic layer caused by the
main separatrix. In Fig.~\ref{SP}(b) the same section is shown for
$E=0.511758756$. We can observe a large  number of islands due to
resonances and a larger chaotic area near the main separatrix. We
also point out the gap in the KAM tori which occurs around $a=0$ and
$p_a=\pm 0.7$, approximately. We shall shortly see how they are filled.

\begin{figure}[t]
\vskip +0.2cm
\begin{center}
\epsfig{file=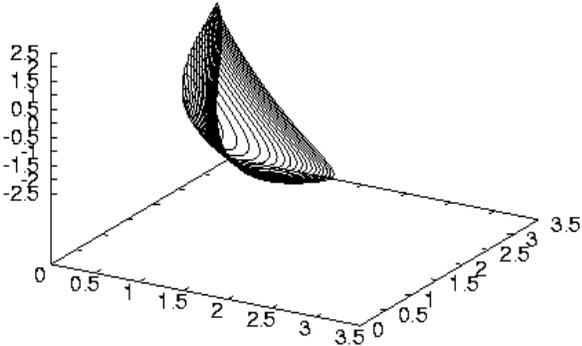,width=\columnwidth,height=5cm}
\end{center}
\caption{Periodic orbits, projection $(a,p_a,p_{\phi})$, for increasing values
of their energies, around the fixed  point $(a,\phi)=(1.0,1.0)$.}
\label{concha}
\end{figure}

\begin{figure}[t]
\begin{center}
\epsfig{file=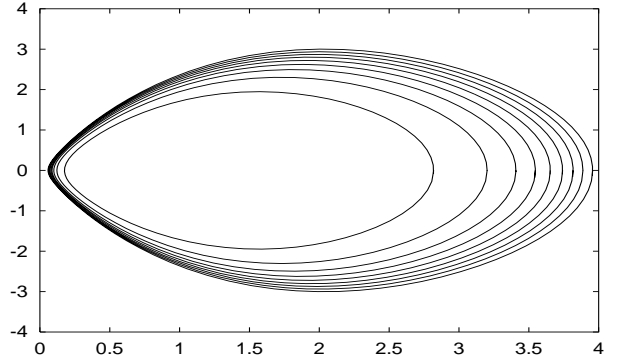,width=\columnwidth,height=5cm}
\end{center}
\caption{The $(a,p_a)$ projection of some very large orbits. We believe that
they go asymptotically to collapse.}
\label{conchap}
\end{figure}

In Fig.~\ref{concha} we can see  part of the center manifold which
contains the fixed point $B_{++}$. From each of these periodic orbits
the unstable cylinder emanates, and the stable one has these orbits
as their asymptotic limit as $\eta \rightarrow \infty$. As mentioned
above, these orbits have been obtained by numerical continuation.
Figure~\ref{conchap} shows the projection of large periodic orbits onto the
$(a,p_a)$ plane. In fact, they get larger and larger
 and seem to approach the collapse as the energy increases.

\begin{figure}[t]
\begin{center}
\epsfig{file=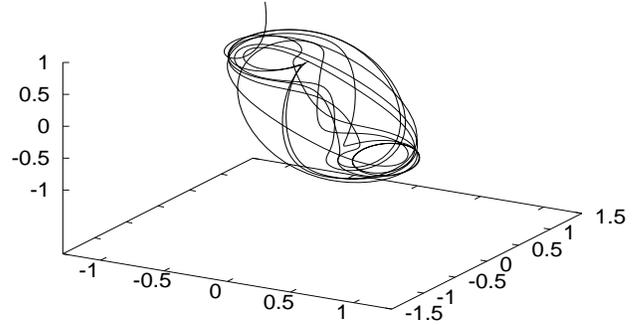,width=\columnwidth,height=5cm}
\end{center}
\caption{An orbit  started on the unstable manifold of the $p.o._{B_{+-}}$
approaching the $p.o._{B_{-+}}$ (front-right) making seven universe cycles on
the process and finally escaping though the unstable manifold of
$p.o._{B_{-+}}$.}
\label{EX}
\end{figure}

From the fixed point of periodic orbits  in the Poincar\'e section we can
globalize the stable and unstable manifolds of the corresponding periodic orbit.
In Fig.~\ref{EX} we show a trajectory bouncing seven times between the the
neighborhood of a  periodic orbit (p.o.) around $B_{-+}$ and that of the p.o.
around $B_{+-}$, before escaping through the unstable manifold of the former.
Orbits like that are necessary to numerically find whether the manifolds
meet tangentially (integrable case) or if they meet at  heteroclinic or
homoclinic connections.

\begin{figure}[t]
\begin{center}
\epsfig{file=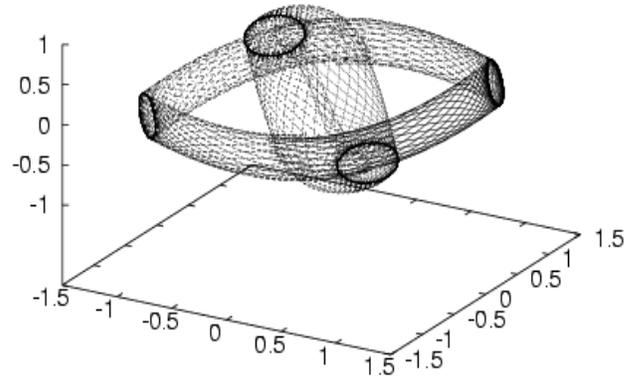,width=\columnwidth,height=6.0cm}
\end{center}
\caption{We can see the four periodic orbits around the fixed points
 $B_i$ and their respective (un)stable  manifolds. They intercept the
 Poincar\'e section $\phi=0$ $(\dot \phi >0)$, which is a symmetry
plane. We  stress that the cylinders which come or go to infinity are
not shown to avoid  cluttering.}
\label {cilindros}
\end{figure}

\begin{figure}[t]
\begin{center}
\vskip -.5cm
\epsfig{file=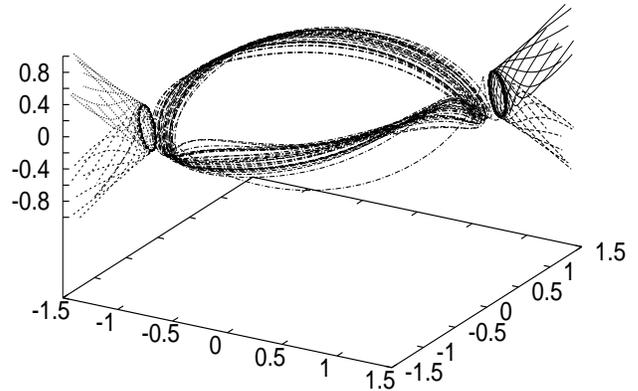,width=\columnwidth,height=6.0cm}
\end{center}
\vskip -.3 cm
\caption{The periodic orbits around $B_{++}$ and $B_{--}$ with
the corresponding unstable and stable cylinders which escape
to  (or arrive from) infinity. The orbit shown starts on the unstable
manifod near $B_{++}$ and escapes through the unstable manifold
of $B_{--}$ after 29 universe cycles.}
\label{cilindros2}
\end{figure}

\begin{figure}[t]
\begin{center}
\vskip -.5cm
\epsfig{file=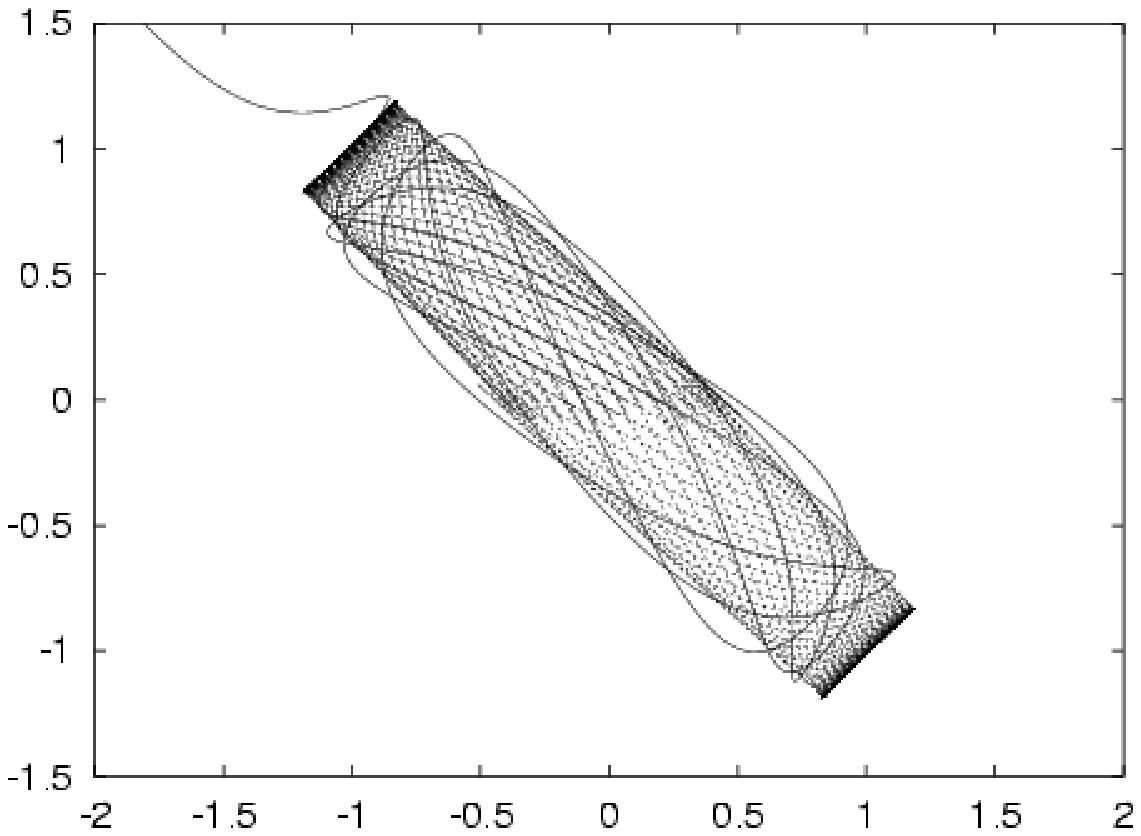,width=4.22cm,height=4.5cm}
\hskip -.1cm
\epsfig{file=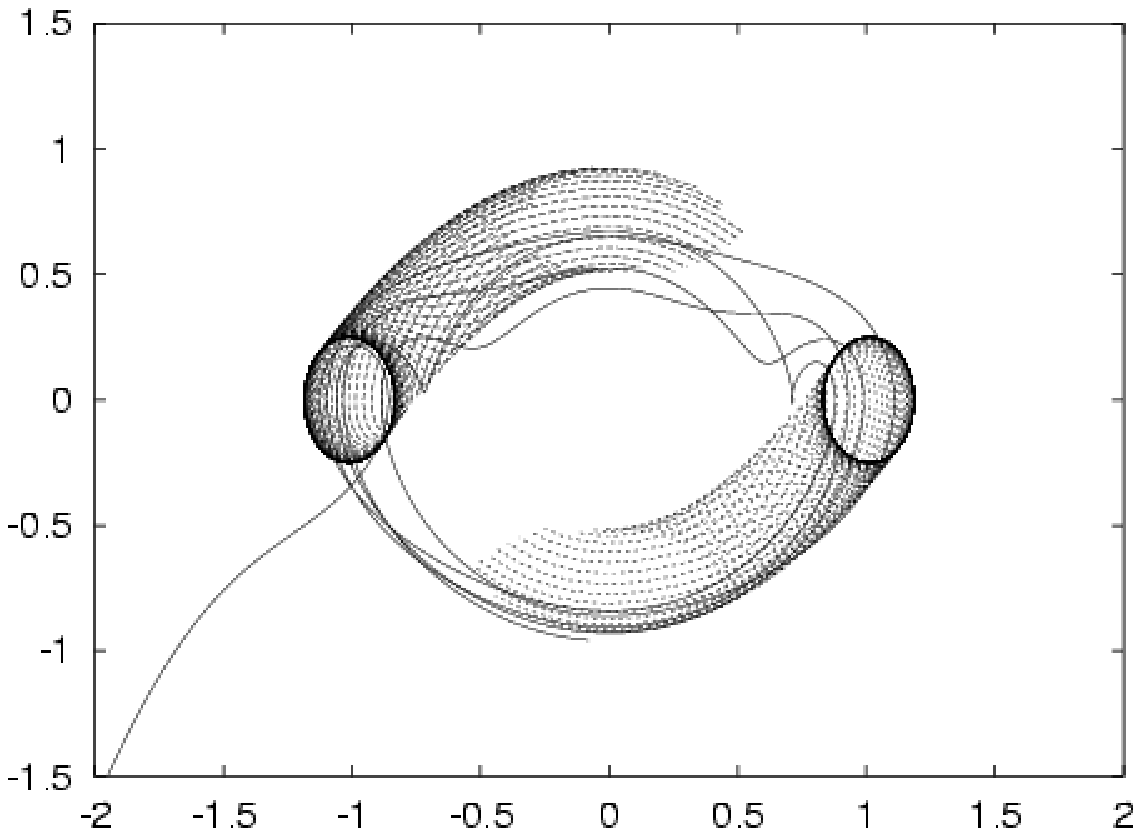,width=4.22cm,height=4.5cm}
\end{center}
\caption{In both figures we show part of the stable and the unstable manifolds
to the periodic orbits at $E=0.56255168$. Left: The projection onto
configuration space of a trajectory starting near the crossing of the unstable
manifold  of $B_{+-}$ with the stable one from $B_{-+}$. After four
cycles it finally escapes to infinity near $B_{-+}$. Right: the projection
onto the $(a,pa)$ plane.}
\label {ciclos}
\end{figure}

\begin{figure}[t]
\begin{center}
\vskip -1.0cm
\epsfig{file=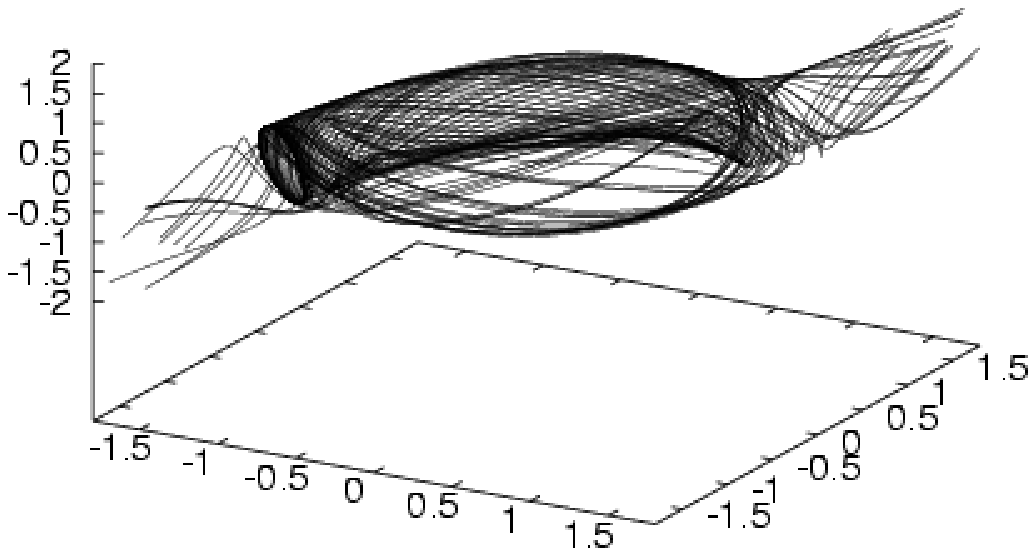,width=4.25cm,height=4.5cm}
\hskip -.2cm
\epsfig{file=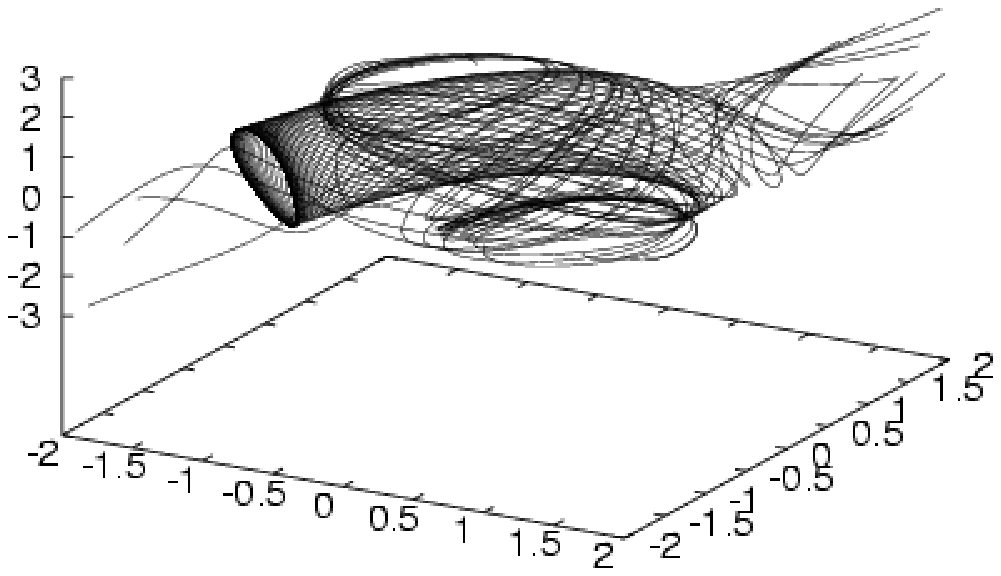,width=4.25cm,height=4.5cm}
\end{center}
\caption{Left: Forty escaping or recolapsing  trajectories
$(a,\phi,p_a)$ starting on the unstable manifold with $a<0$ (near the
$p.o._{B_{-+}}$) for $E=0.7307128$. Right: the same for $E=1.09004$. All
orbits were allowed four passages through the Poincar\'e section.}
\label{colsca}
\end{figure}

Part of the globalized cylinders are shown in Fig.~\ref{cilindros}, where we
show a three-dimensional picture in the  variables $(a,\phi,p_a)$. We have
stopped the iteration at  the first intersection with the Poincar\'e section
$\phi=0$ ($\dot \phi>0$).  Note that we have not shown the portions of the
unstable and stable manifolds which escape to (or come from)  infinity to avoid
cluttering the picture. For this same reason we exhibit just a few orbits on
each manifold. We  can  clearly see the four intersections  of the cylinders:
the unstable (stable) cylinder to  $p.o._{B_{+ +}}$ with the stable (unstable)
one to $p.o._{B_{- -}};$  the same for $p.o._{B_{+ -}}$ with $p.o._{B_{- +}}$
--- where $p.o._{B_i}$ stands for periodic orbit around the fixed point $B_i$.
We note that only two of the four intersections belong to the same Poincar\'e
section. To see the other pair, one should just take the oposite sign ($\dot
\phi<0$).
	As is well known, if the stable and unstable manifolds intersect each other
once, they must intersect each other an infinite number of times. Since each
such crossing is invariant, their intersections must stay on those cylinders
upon both forward and backward iterations of the dynamics. Therefore, they are
biasymptotic to the p.o.'s. The Poincar\'e map of the heteroclinically
connecting trajectories initially evolving, say,  on the unstable manifold, will
then follow the stable one, and from then on it will approach the p.o.
around the opposite (w.r.t. the origin) fixed point as $\eta \rightarrow
\infty$. On the other hand, as $\eta \rightarrow -\infty$ the trajectory
travels towards the original fixed point (of the original p.o.).
However, the heteroclinic trajectory  itself  has  to pass through all these
points, making then an infinite number of universe cycles and taking an infinite
time to do so. To describe the dynamics of a heteroclinic Poincar\'e iterate is
very difficult, let alone four heteroclinic orbits traversing in the
same region. However, we can show that trajectories in  the neighborhood of
the cylinders may, after some universe cycles, go either to collapse or escape
to infinity, strongly depending, in an unpredictable way, on their initial
conditions. In Fig.~\ref{cilindros2} we exhibit an orbit starting on the
unstable manifold of $p.o._{B_{++}}$ and performing 29 universe cycles before it
escapes through the unstable manifold of $p.o._{B_{--}}$ (note the darker line
on the left bottom manifold).
In Fig.~\ref{ciclos} we show another example,  where an orbit makes four
universe cycles and finaly escapes to infinity (this part of cylinders are not
shown to avoid cluttering). In Figures~\ref{colsca}(a) and \ref{colsca}b, for
$E=0.7307128$ and $E=1.09004$ respectively, we show fourty orbits starting on
the unstable manifold of $p.o._{B_{--}}$ and  either recollapsing or escaping
towards $a \rightarrow \infty$. The presence of the  heteroclinic connections,
as seen above, unquestionably assures that such a system is chaotic.

	Before going to the next aspect of the dynamics, we shall analyse another piece
of infomation that can be extracted from Figures~\ref{colsca}(a) and 
\ref{colsca}(b): the lower energy orbits escape {\bf only} through the
symmetrical point with respect to their initial conditions, while some
orbits at higher energies  escape also through the transverse connections. This
can be developed into a tool for reckoning the escape rate noncollapsing
universes (i.e., those which do not cross $a=0$ or $\dot a>0$).

\begin{figure}[t]
\begin{center}
\vskip -0.5cm
\epsfig{file=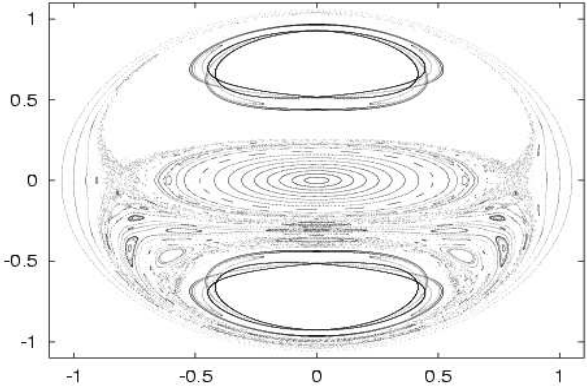,width=\columnwidth,height=5.5cm}
\end{center}
\caption{Poincar\'e section  of the cylinders  and KAM tori for
 $E=0.562551684$, same as Fig.~\ref{hetcon2}.  Universes which before were bound
 to do only cycles can now find a way to scape to infinity. }
\label{destr1}
\end{figure}

 Another interesting feature is how the heteroclinic cuts blend
with cuts of KAM tori and irregular regions.
\begin{figure}[t]
\begin{center}
\epsfig{file=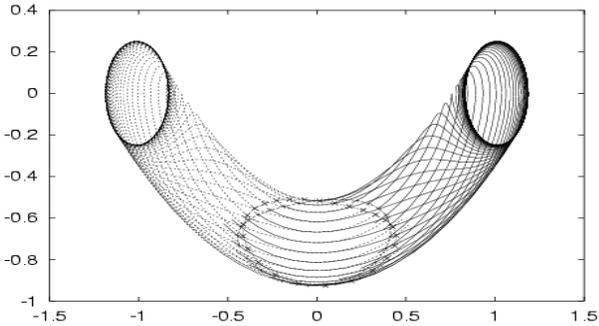,width=\columnwidth,height=4.5 cm}
\end{center}
\caption{Poincar\'e section and projection of the unstable
manifold of $p.o._{B_{++}}$ and the stable manifold of $p.o._{B_{--}}$.
The four intersections of the Poincar\'e section is enhanced with
help of lines.}
\label{hetcon0}
\end{figure}
\begin{figure}[t]
\begin{center}
\epsfig{file=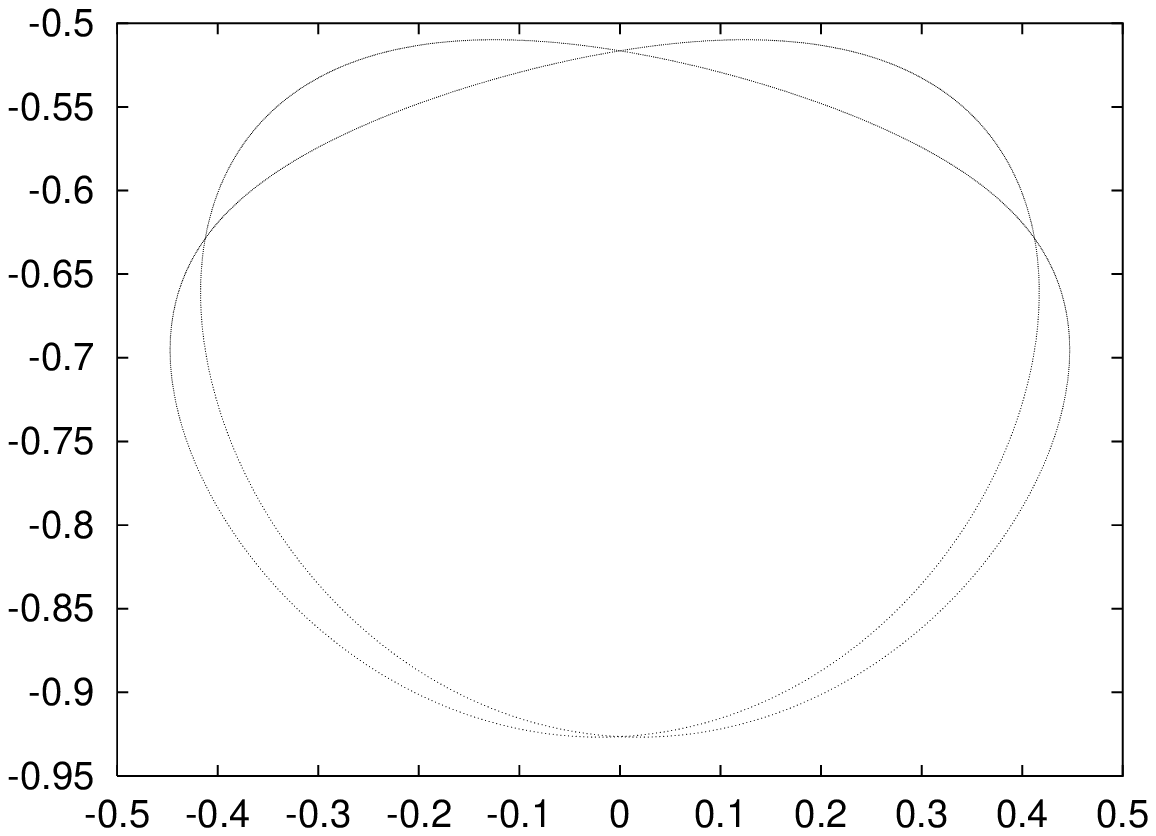,width=4.22cm,height=4.5cm}
\hskip -.2 cm
\epsfig{file=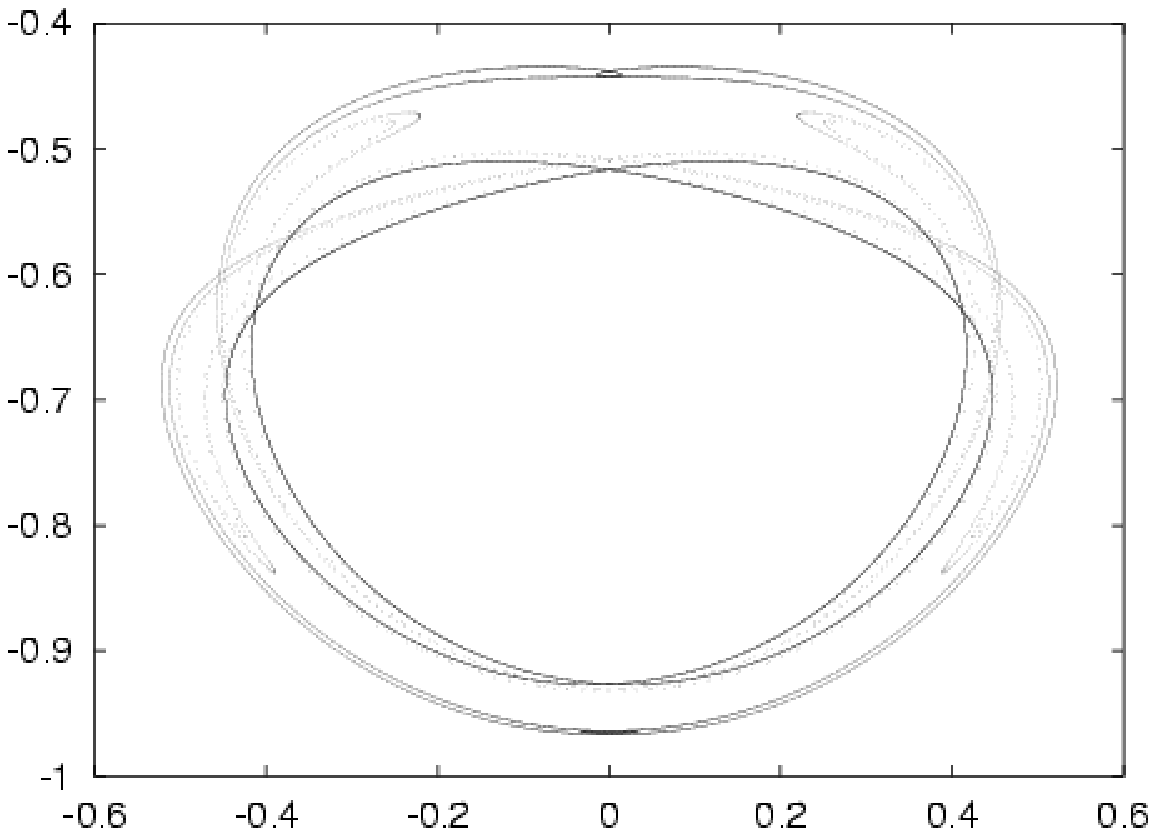,width=4.22cm,height=4.5cm}
\end{center}
\caption{{\bf (a)} Same as Fig.~\ref{hetcon0} with 1000 initial
contiditions  showing the first cut of the unstable manifold of $p.o._{B_{++}}$
and the stable one of $p.o._{B_{--}}$; {\bf (b)} Same as (a),  adding the
iterates  of the second passage through the Poincar\'e section (10000 initial
conditions).}
\label{hetcon1}
\end{figure}
 Figure~\ref{hetcon0} shows the unstable manifold of $p.o._{B_{++}}$ and
the stable manifold of $p.o._{B_{--}}$  projected in the plane of the
Poincar\'e section $(a,p_a)$. In this same picture we show the
points (crosses) where those manifolds cut the section. The lines joining them
are drawn just to help visualization. In   Fig.~\ref{hetcon1}(a) we show the
first two cuts (one from each manifold) with a large number (1000) of initial
conditions so that the section looks like  a continuous curve. The existence of
four heteroclinic  connections can be easily seen.  Figure~\ref{hetcon1}(b)
shows two consecutive cuts of each manifold with an even larger (10000) number
of points. This increase is necessary because the initial cuts evolve to highly
convoluted ones due to the folding and stretching of the cylinders.  To follow
this picture in detail is beyond the scope of this work.
We now superpose the Figs.~\ref{SP}(b) and \ref{hetcon1}  and find that
the heteroclinic cuts are embeded among the tori in the upper and
lower part of their Poincar\'e section, exactly in the gaps that we
showed in Fig.~\ref{SP}b. Then, besides the breaking of KAM
tori due to the resonances of their frequencies, for $E>E_B=0.5$,
 they are also eroded by the unstable and stable manifolds of the periodic
orbits with the same energy. This can be seen clearly in
Figure~\ref{destr1}. For the sake of clearness, we show the complete
lower part ($p_a<0$) and only the heteroclinic cuts for the upper
part (due to the cylinders of $p.o._{B_{+-}}$ and $p.o._{B_{+-}}$).
 They merge nicely with the KAM sections and are another source
of destruction of regular behaviour.

\begin{figure}[t]
\begin{center}
\epsfig{file=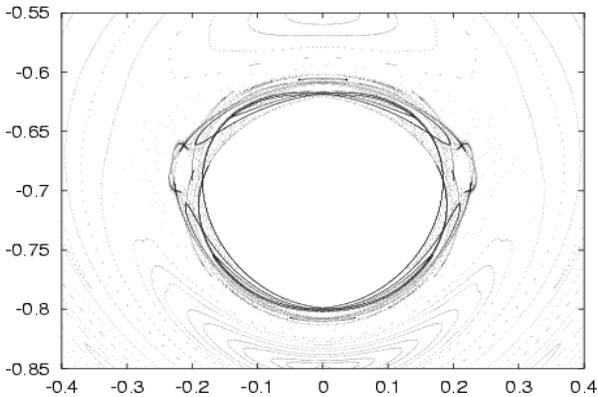,width=\columnwidth,height=5.5cm}
\end{center}
\caption{Detail of Poincar\'e section  of the cylinders  and KAM
tori for $E=0.511758756$. Note, at top and bottom, trapped islands among the
 five cuts of the cylinders, and the gaps at left and right.}
\label{destr2}
\end{figure}

\begin{figure}[t]
\begin{center}
\vskip -0.5cm
\epsfig{file=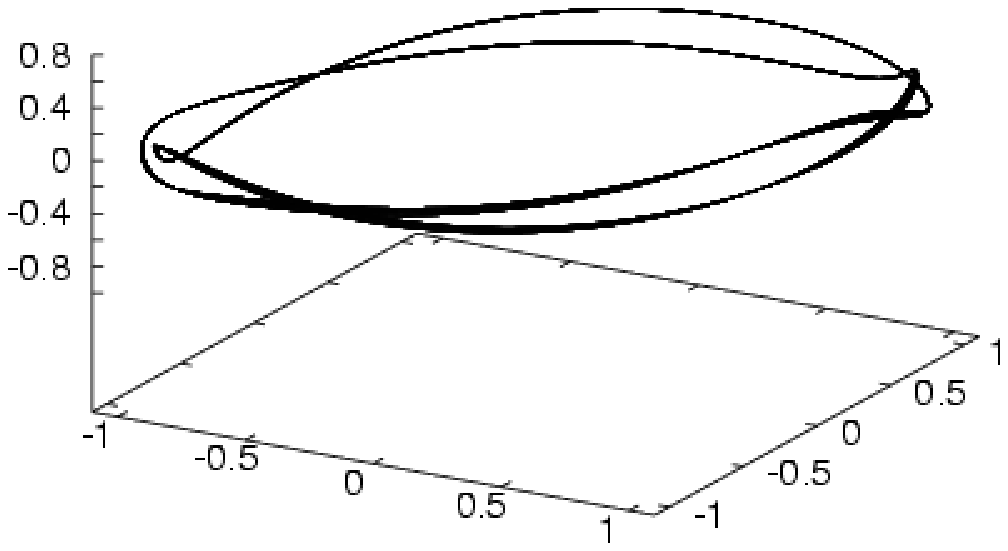,width=4.22cm,height=4cm}
\hskip -.2cm
\epsfig{file=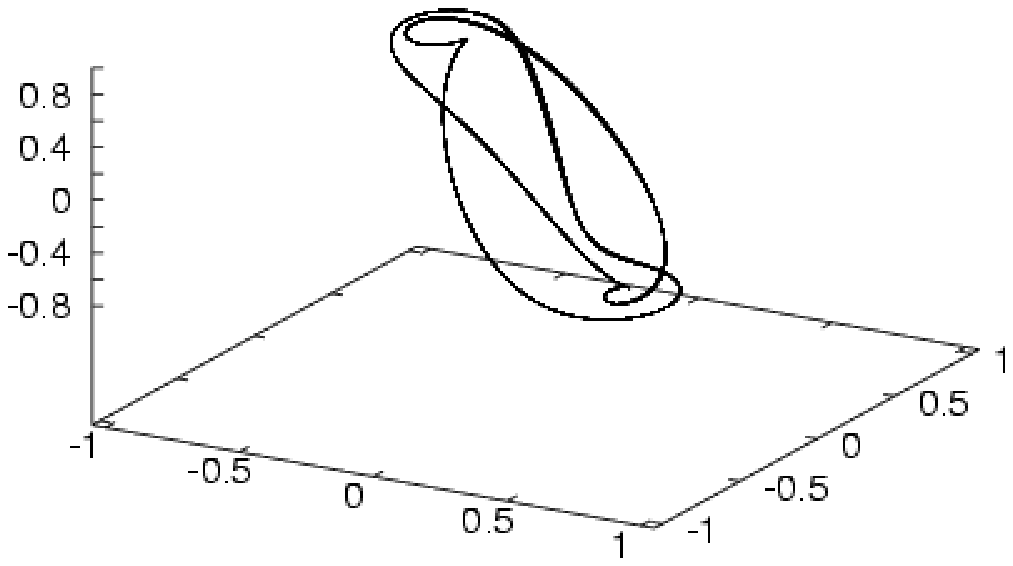,width=4.24cm,height=4cm}
\end{center}
\caption{Left: The projection onto $(a,\phi,p_a)$ of a trajectory lodged among
the cuts at top and bottom of the previous figure. Right: the same projection of
a KAM torus  which in the section $\phi<0$ would occupy the gaps on both sides
of Fig.~\ref{destr2} (see the text for explanation).}
\label {smile}
\end{figure}

 In Fig.~\ref{destr2} we show a more detailed view of the merging for
the energy $E=0.511758762$. There are islands lodged among the cuts of the
cylinders (top and bottom): they manage to avoid escaping and are bound to
recollapse. We show a particular trajectory corresponding to one of these
islands in Fig.~\ref{smile}(a).  The empty regions at both right and left of
Figure~\ref{destr2} belong to the  section with oposite crossing sign.  The
orbit shown in Fig.~\ref{smile}(b)  confirms this hypothesis: they traverse
from the neighborhood of $B_{+-}$ and $B_{-+}$.  It is interesting how these
orbits avoid escaping due to a resonance between $\dot a$ and $\dot \phi$.  Note
that both orbits get very near the unstable periodic orbits.

\begin{figure}[t]
\begin{center}
\epsfig{file=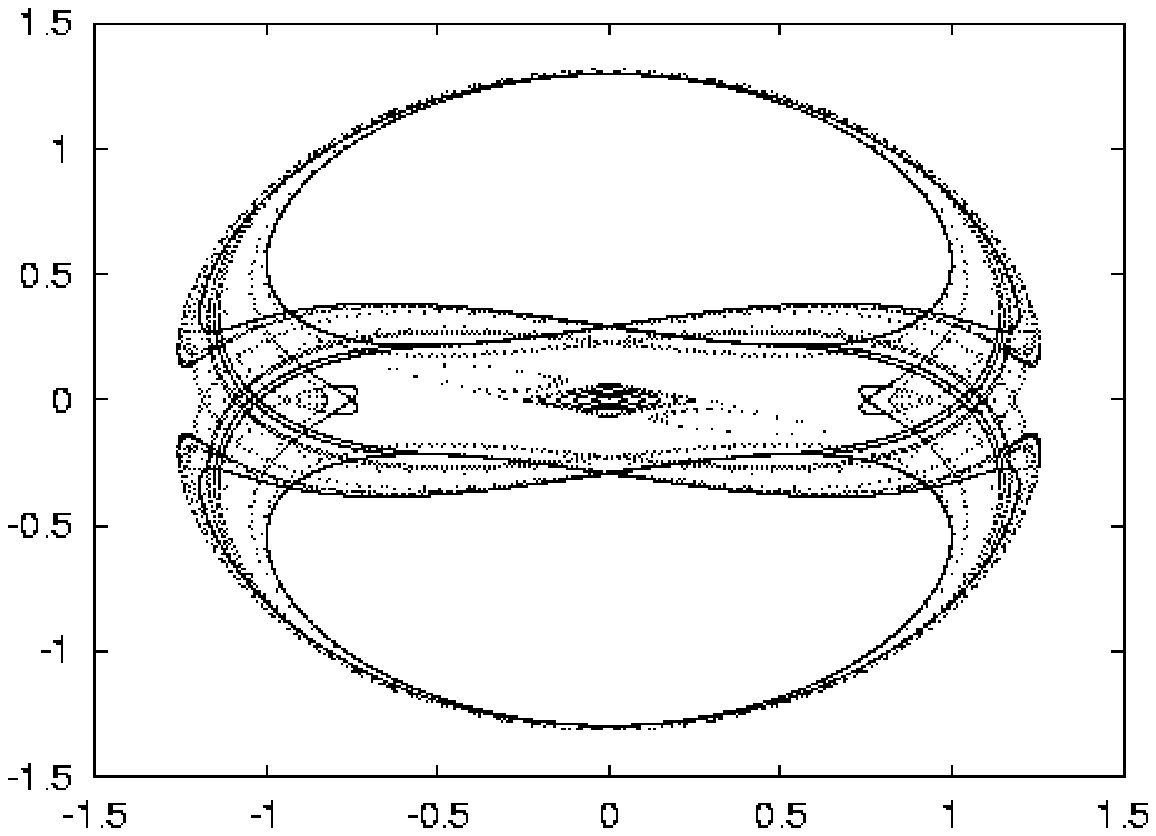,width=4.1cm,height=5cm}
\epsfig{file=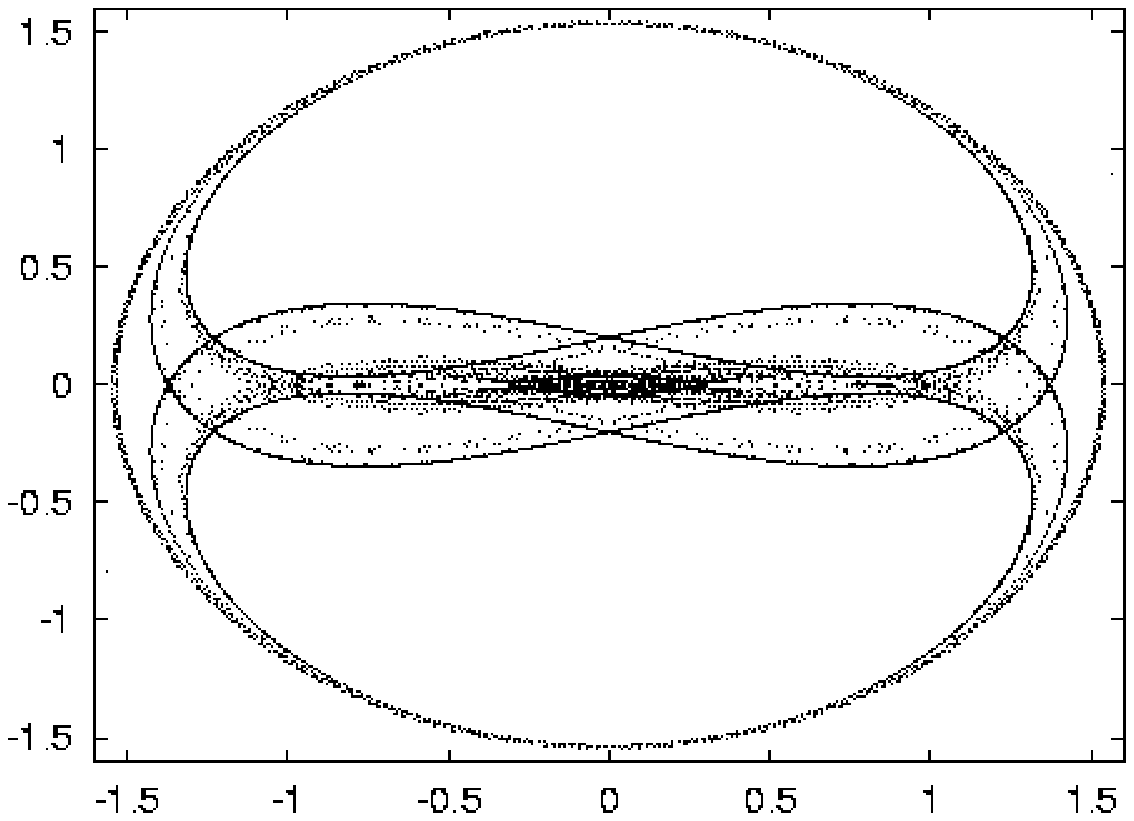,width=4.1cm,height=5cm}
\epsfig{file=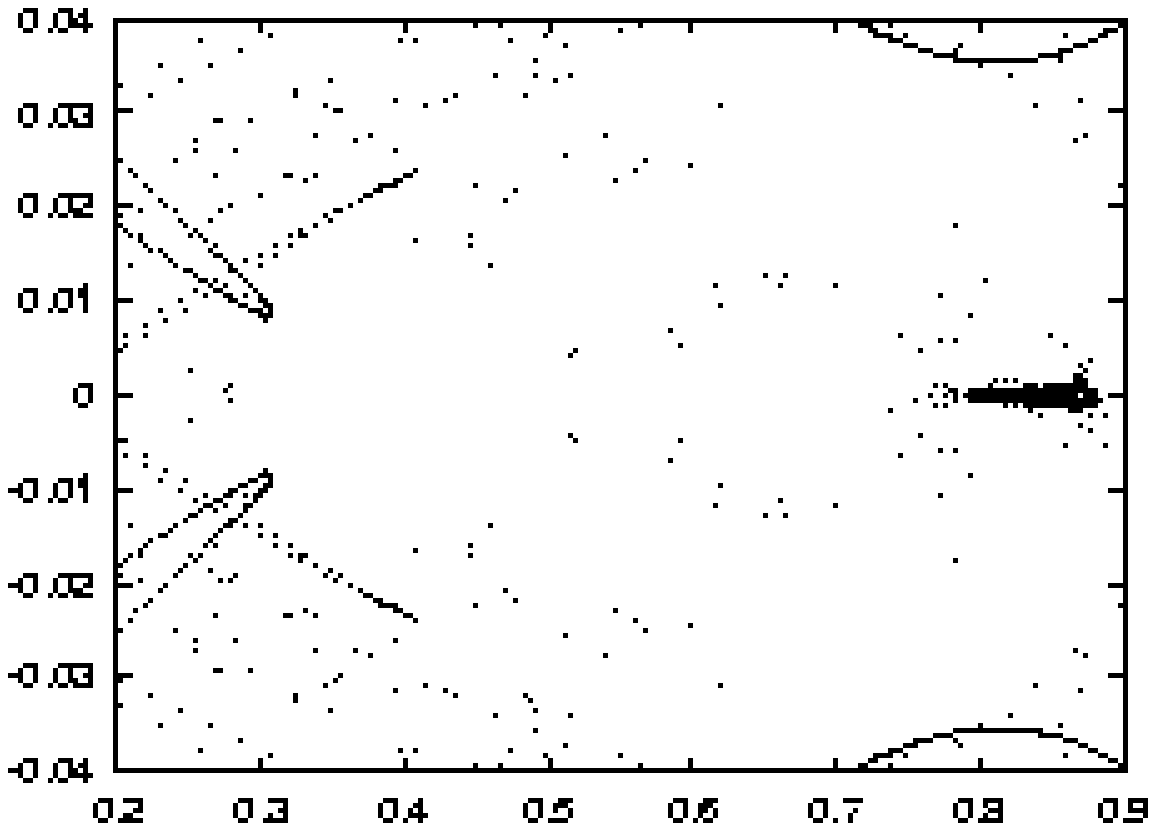,width=\columnwidth,height=7cm}
\epsfig{file=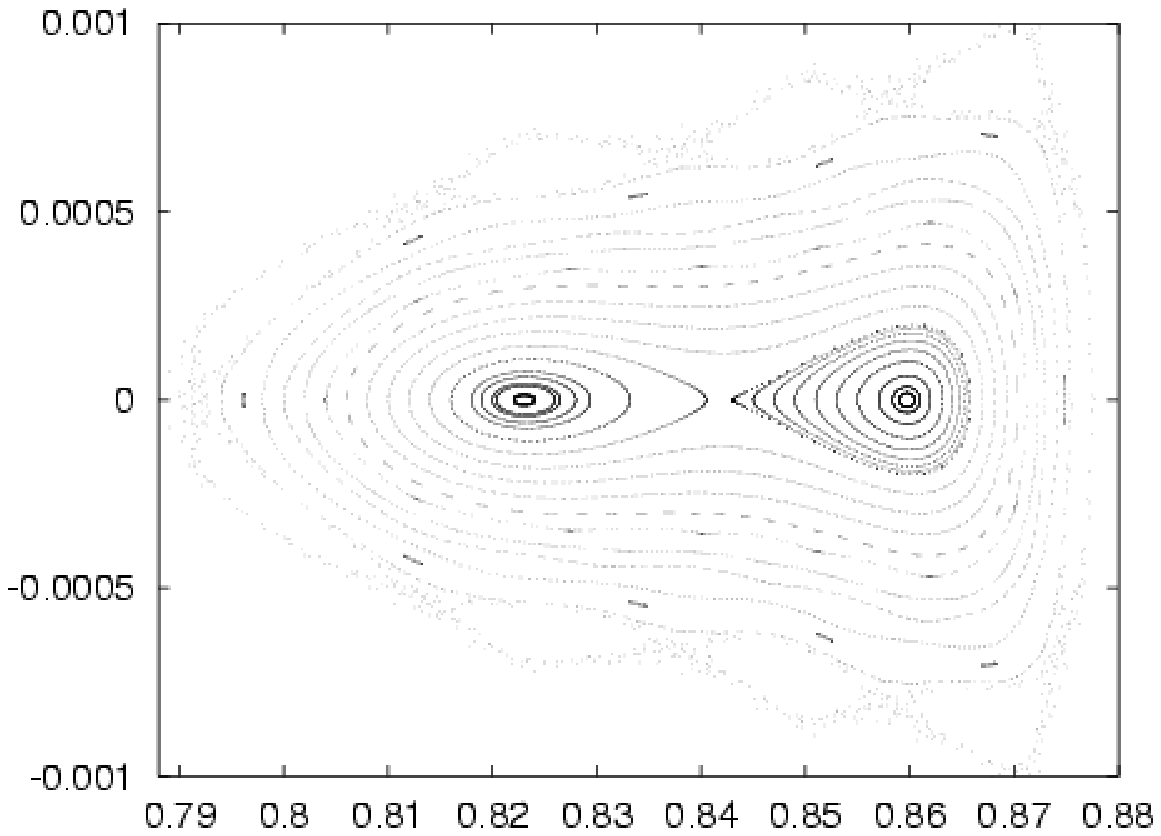,width=\columnwidth,height=6cm}
\end{center}
\caption{Two iterates of the stable and unstable manifolds
to a large periodic orbit of $B_{++}$ for $E=0.886408856$ and
$E=1.19549093$, respectively. In the bottom, we zoom in the latter
to show a detail of the remnant tori structure squeezed
between the (un)stable manifolds. Note the axis range of another zoom into
what looks like a black spot in the previous magnification.}
\label{hetcon2}
\end{figure}

We show in Fig.~\ref{hetcon2}a-b how the KAM tori are squeezed near the
trivial fixed point $(0,0)$ as the energy increases,  and  are finally
completely destroyed by the (un)stable manifolds.  In Fig.~\ref{hetcon2}(a) we
show the   section for $E=0.886408856$ where a tiny region near the
trivial fixed point still presents KAM tori.  As the cylinders expand, the
trival fixed point bifurcates.  This bifurcation also
happens in the real counterpart, as expected; we will postpone this dicussion
until the next section.

\begin{figure}[t]
\begin{center}
\epsfig{file=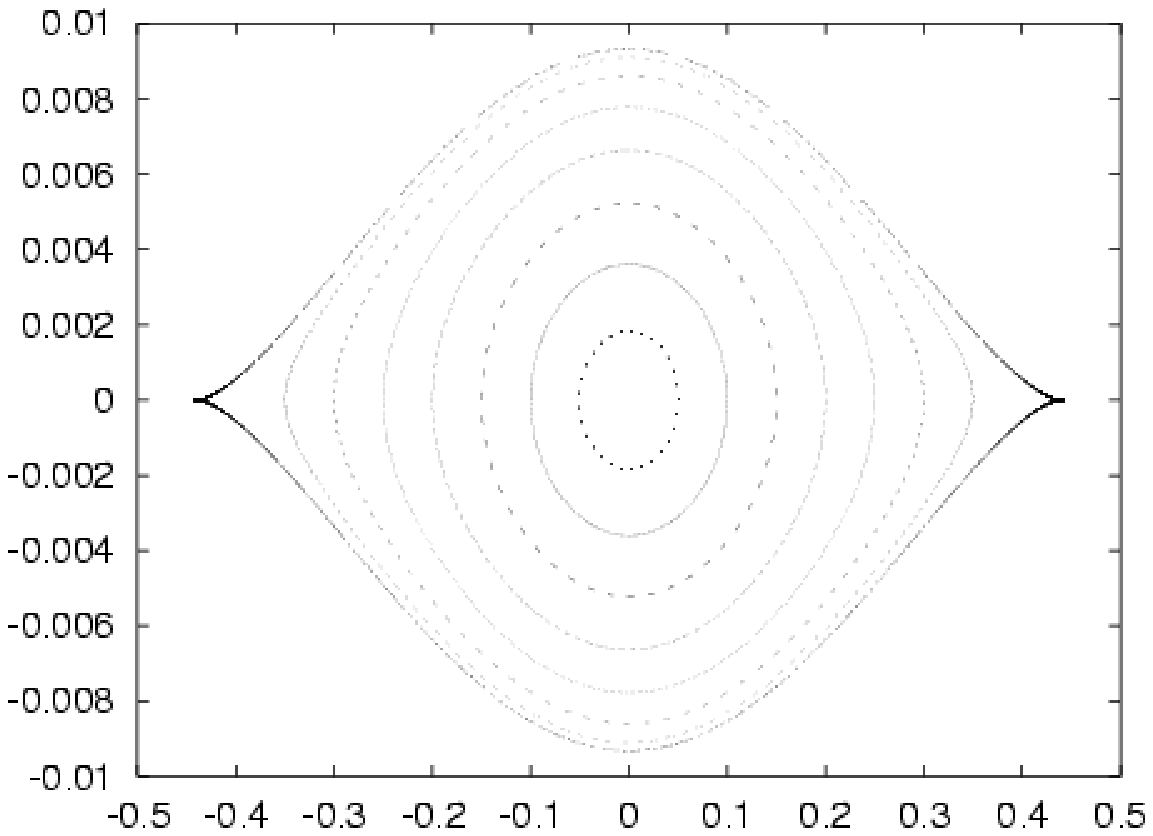,width=\columnwidth,height=5cm}
\epsfig{file=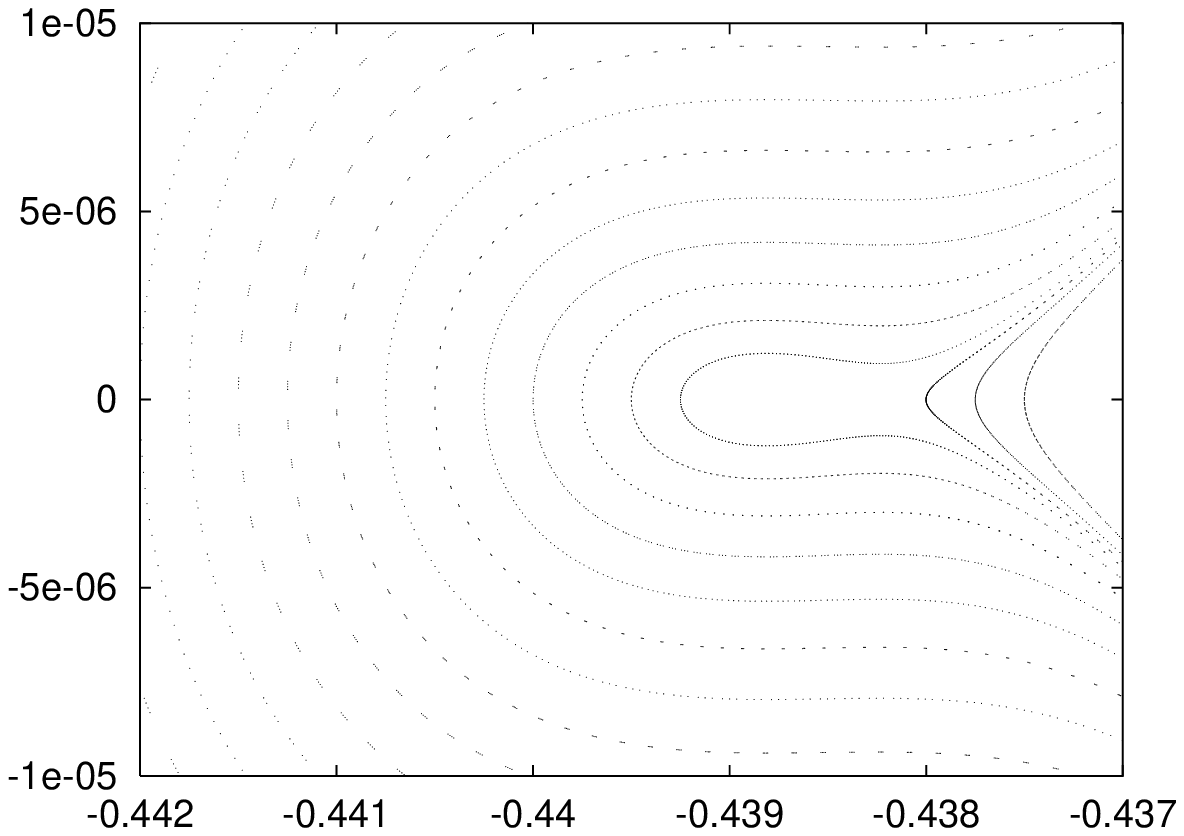,width=4.1cm,height=5cm}
\epsfig{file=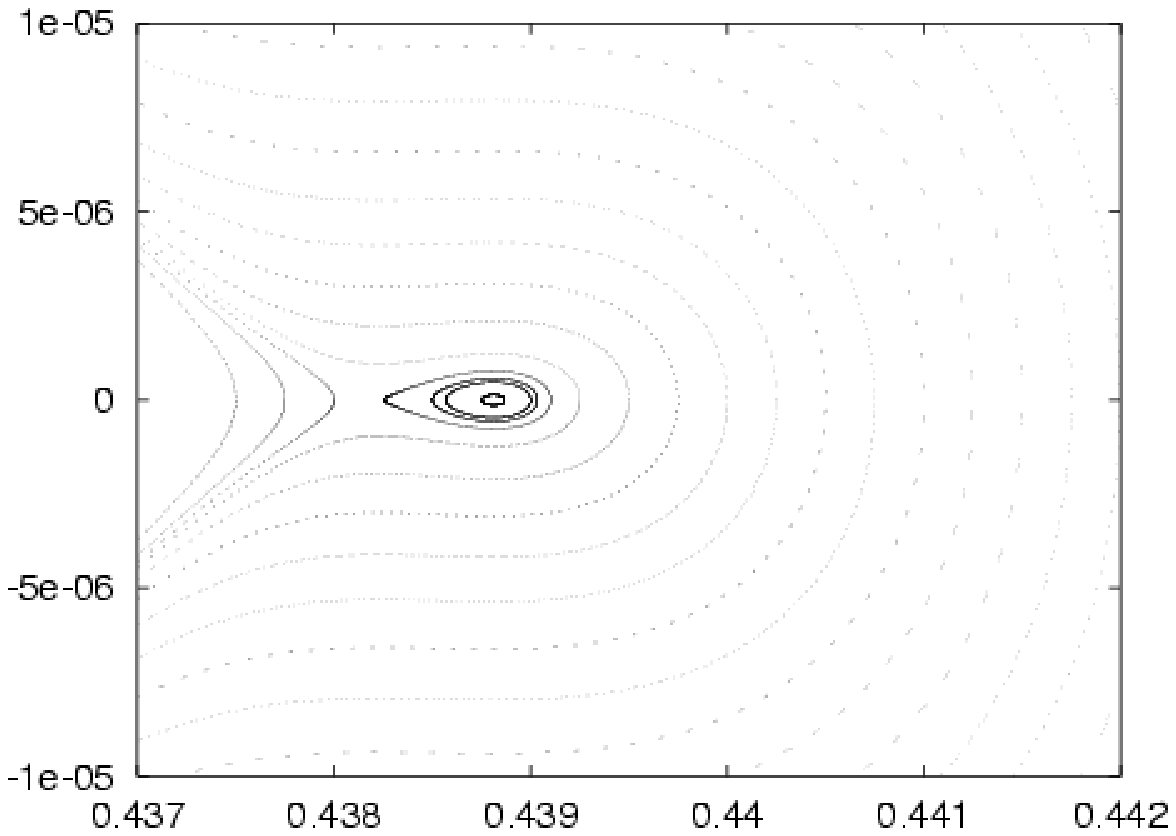,width=4.1cm,height=5cm}
\end{center}
\caption{The bifurcation at $E=1.15769143$ where the first noncollapsing
universes are born. The KAM structure shown in the previous figure originates
from this bifurcation. It has been pushed away by the expanding (un)stable
manifolds.}
\label{abif}
\end{figure}

Another interesting aspect of the imaginary dynamics is a
$2$:$1$ bifurcation at $E = 1.157691425$. This kind of bifurcation is
typical of  mappings which are nearly integrable and, at some distance of the
equilibrium point,  looses its  twist character (see Ref.~\cite{stuchi2} for
details). In our case, though, it is more important to stress that this
bifurcation creates a region  in which tori starting at $a>0$ ($a<0$) stay at
the same side of the $a=0$ axis. These noncollapsing universes are shown in
Figs.~\ref{abif}. Its bottom right panel shows the small
island where the universes are confined to. Note that in the
bottom left panel (same figure) the corresponding region is empty since we have
given initial conditions with $a>0$. That empty region can only be filled with
iterates of trajectories starting at $a<0$, in that range.

\begin{figure}[t]
\vspace{-.5cm}
\begin{center}
\epsfig{file=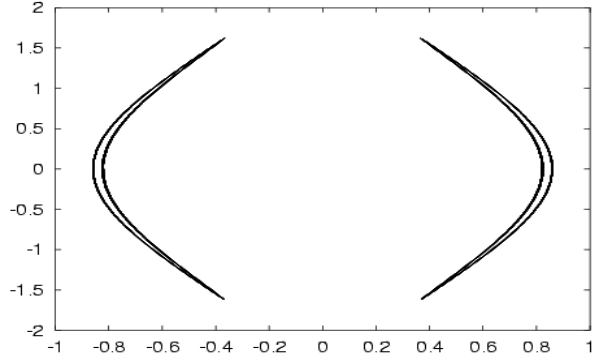,width=\columnwidth,height=5cm}
\end{center}
\caption{Projection onto the configuration space of two
tori which stay at the same side of the axis $a=0$. These orbits belong to
the remnant structure shown in Fig. 9(c) at $E=1.195549093$.}
\label{semcol}
\end{figure}

As the energy inceases, this structure grows and is pushed away to higher
mean values of the radius. In Figs.~\ref{hetcon2}(b) and \ref{hetcon2}(c), we
show the KAM structure left. Note in Fig.~\ref{hetcon2}(b) that the intercepting
curves, formed by the iterates of the Poincar\'e section of the cylinders,
squeeze down the amplitude of the momentum $p_a$ of the remnant KAM tori,  and
out from the central region. Another zoom into the black
spot of Fig.~\ref{hetcon2}(c) shows a further bifurcation, where the
stable periodic orbit becomes unstable yielding two stable ones (the central
points in the section).
%
%
For even larger values of the energy, they finally disappear completely and
 the (un)stable manifolds will be the only  remaining invariant surfaces with
the unstable periodic orbits typical of a horse-shoe structure occupying the
whole phase space. This is a mechanism by which orbits born close to the
singularity may find its way either to infinity or to  collapse in
an unpredictible way. Other orbits would be trapped as heteroclinic ones
approaching  one of the stable periodic orbits at $a>0$ having an infinite
period (see Fig.~\ref{semcol}).

\section{Real problem}
\label{real}

\begin{figure}[t]
\begin{center}
\epsfig{file=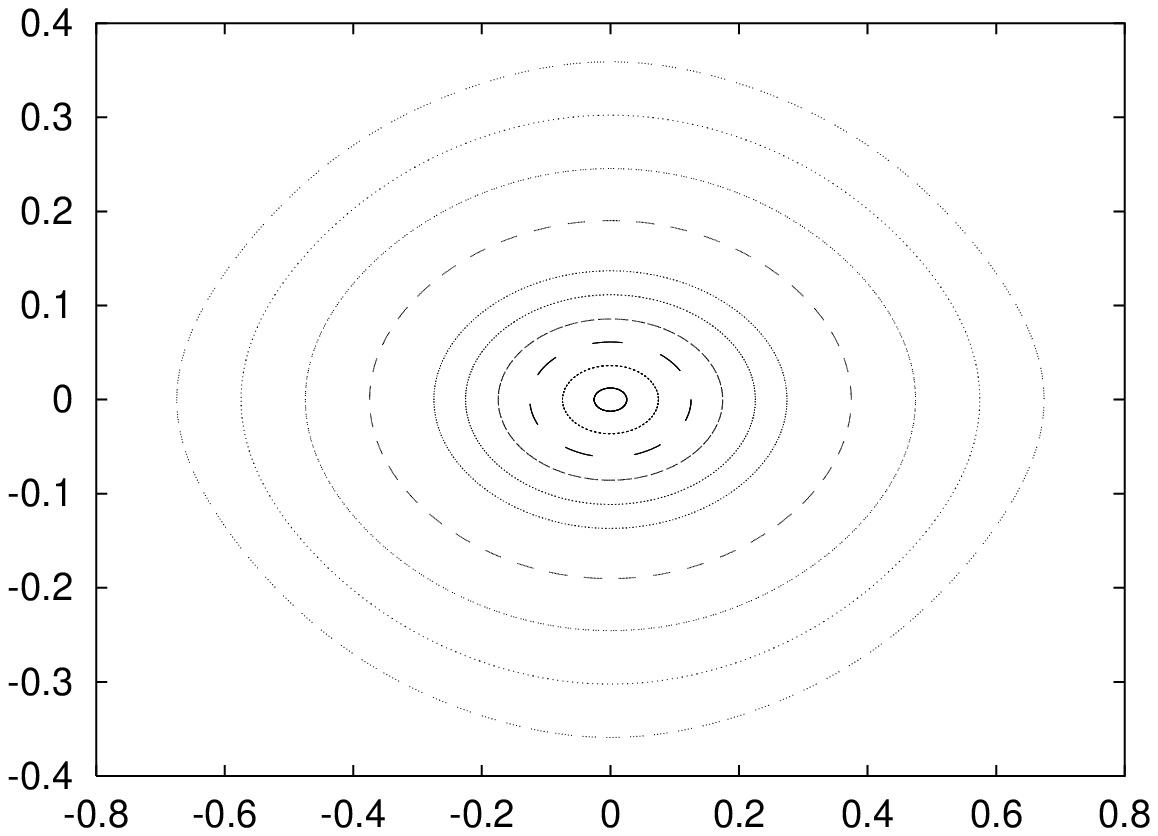,width=4.1cm,height=4cm}
\epsfig{file=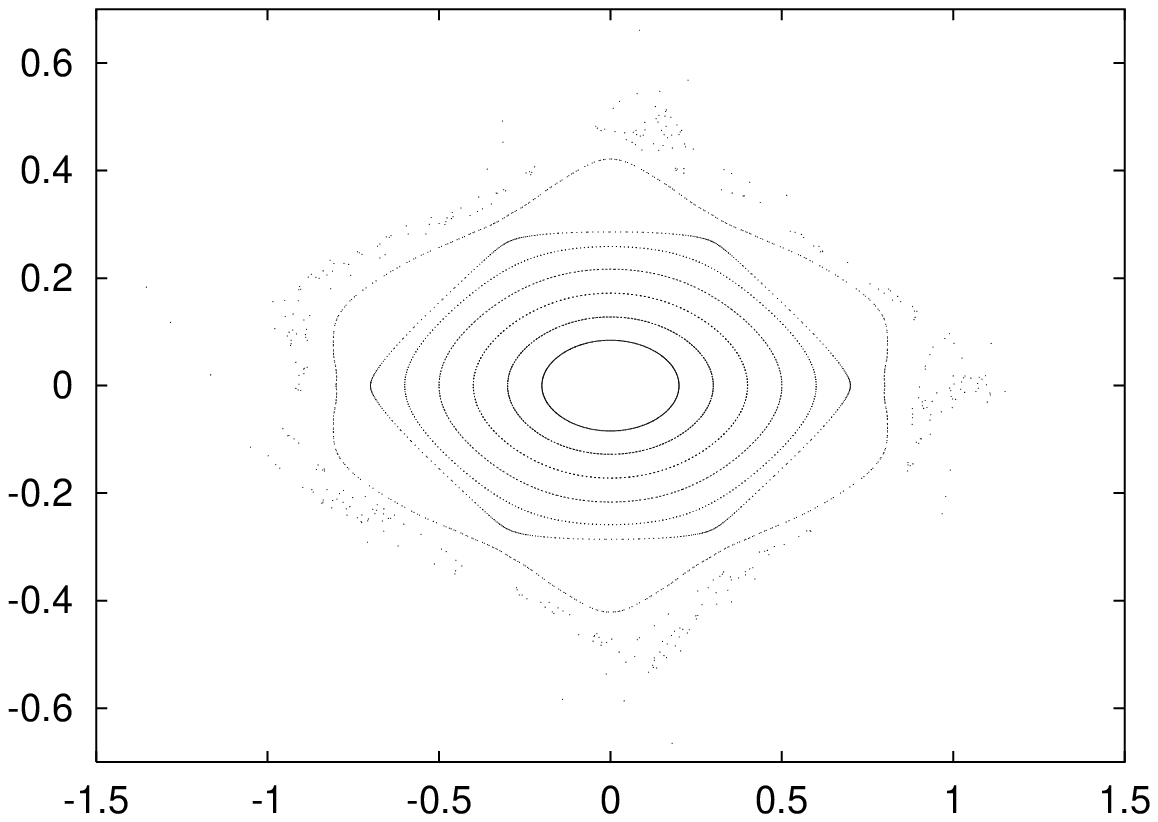,width=4.1cm,height=4cm}\\
\end{center}
\caption{The outer region becomes irregular as the energy increases.
 From left to right: $E=0.0$ and $0.3$. Chaos appears in the outer
region of the phase space when it opens up and trajectories can escape
to infinity performing larger and larger cycles. Note that the
six islands belong to two different  tori; they were drawn with different
initial conditions to enhance this fact.}
\label{show}
\end{figure}

We recall the real Hamiltonian:
\begin{equation}
2H = (p_\phi^2  + \phi^2 ) -  (p^2_a + a^2 ) +
m^2 a^2 \phi^2 + \rho_r = 0.
\end{equation}
Since there is a bifurcation of the shared fixed point at the origin in the
imaginary case, leading to a noncollapsing KAM structure, we now investigate
 the real model numerically. As before, for lower energy, the whole energy
 space is almost integrable, in the sense that eventual resonances (or
 islands) are very thin structures. As the energy increases, bifurcations
 become more frequent and for $E=0.5$ the outer tori are completely destroyed
 while the inner core is kept regular. This is shown in Fig.~\ref{show}, where
 we display some sections through $\phi=0$ ($\dot\phi>0$) for
 increasing values of the energy.

\begin{figure}[t]
\begin{center}
\epsfig{file=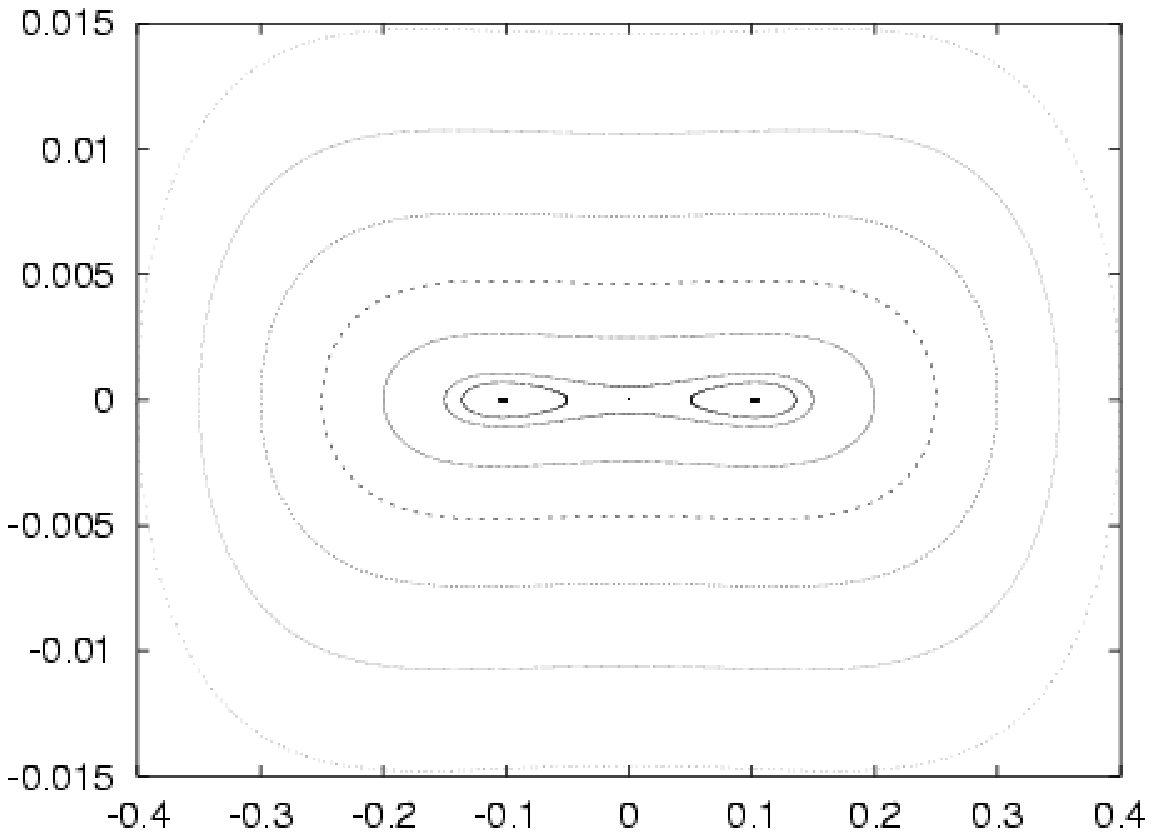,width=4.1cm,height=4cm}
\epsfig{file=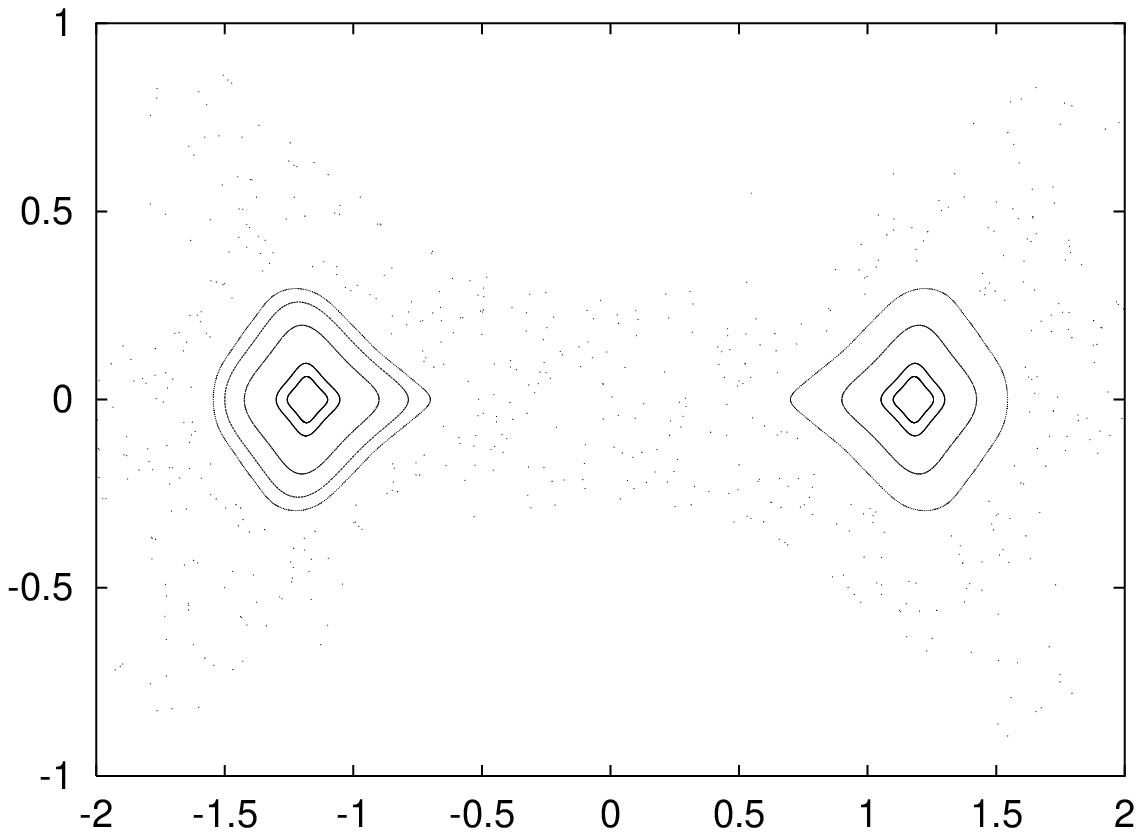,width=4.1cm,height=4cm}
\end{center}
\caption{Two stable periodic orbits are born at the bifurcation of the
trivial fixed point, and the latter becomes unstable. For $E=1.164$ the
region of noncollapsing orbits is still very small. On the right
figure we show $E=1.3$: the outer layer collapses after more 900 Poincar\'e
iterates, shown by crosses.}
\label{bifr}
\end{figure}

\begin{figure}[t]
\begin{center}
\epsfig{file=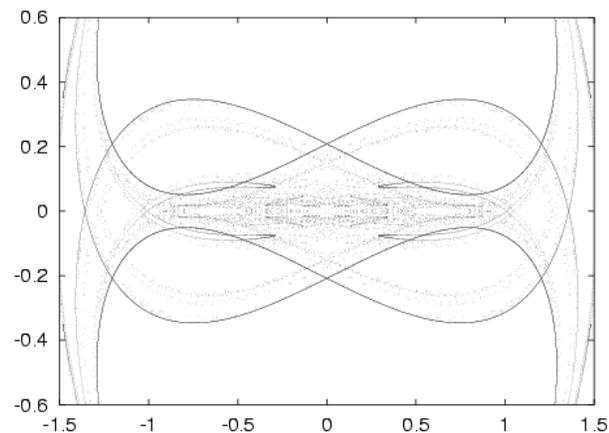,width=\columnwidth,height=6cm}
\end{center}
\caption{We show 20 passages of the stable and unstable manifolds of the
analytical continuation for $E=1.16682669$. Note how the origin is gradually
surrounded by the intercepting loops of these manifolds.}
\label{uscbif}
\end{figure}

For $E\approx 1.16335949$, the trivial fixed point bifurcates from stable to
unstable, giving birth to two stable periodic orbits, as shown in
Fig.~\ref{bifr}. This feature is, of course, shared by the analytical
extension as mentioned in the previous section. In Fig.~\ref{uscbif} we
show 20 iterates of the stable and unstable manifolds. Note how
they form intercepting loops which approach the origin, leaving no room for
regular motion around the trivial fixed point: the real model is simply obliged
to bifurcate. This is the origin of the chaos at large in the analytical
extension.  However, as we shall see, this bifurcation favors some regularity
and noncollapsing KAM structures.

 The newborn periodic stable orbits and surrounding KAM tori do not collapse.
Two of such tori are seen in Fig.~\ref{rpo} below. The orbit shown in the left
of this figure  is very near the newborn stable periodic orbit
at the bifurcation energy.


%

\begin{figure}[t]
\begin{center}
\epsfig{file=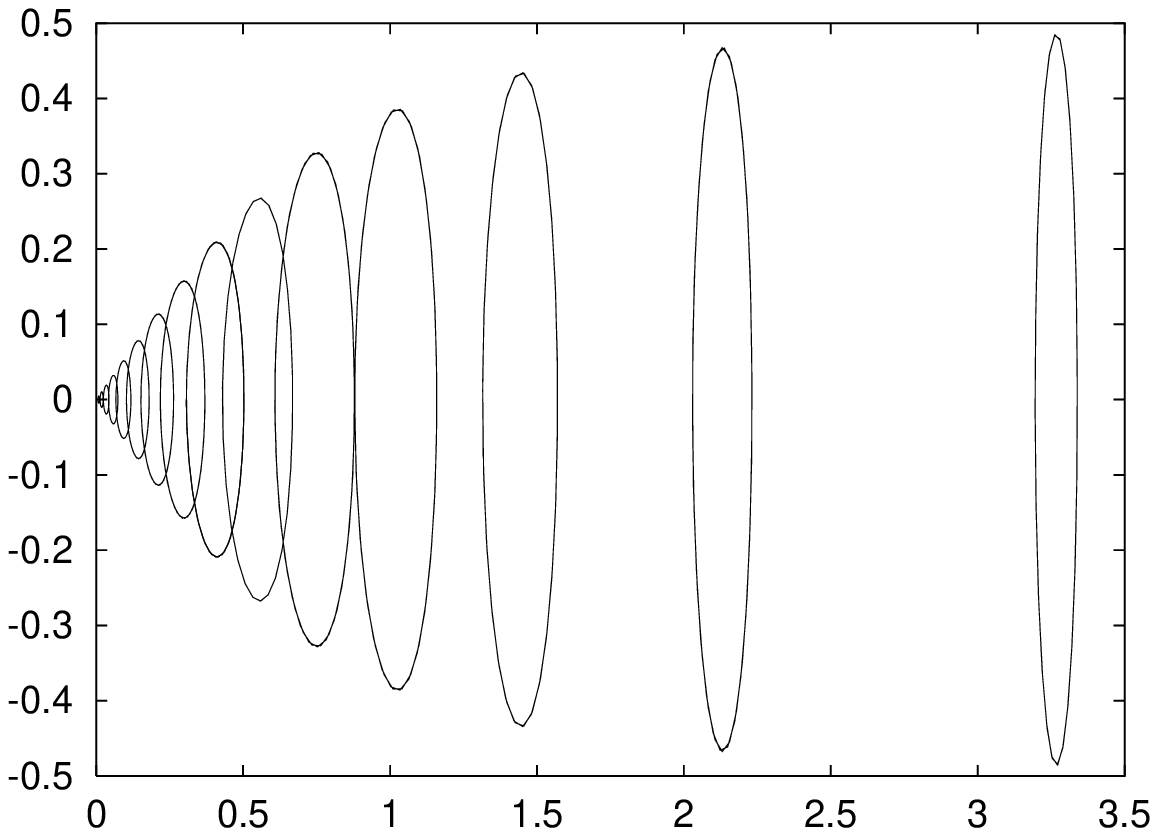,width=4.1cm,height=4.5cm}
\epsfig{file=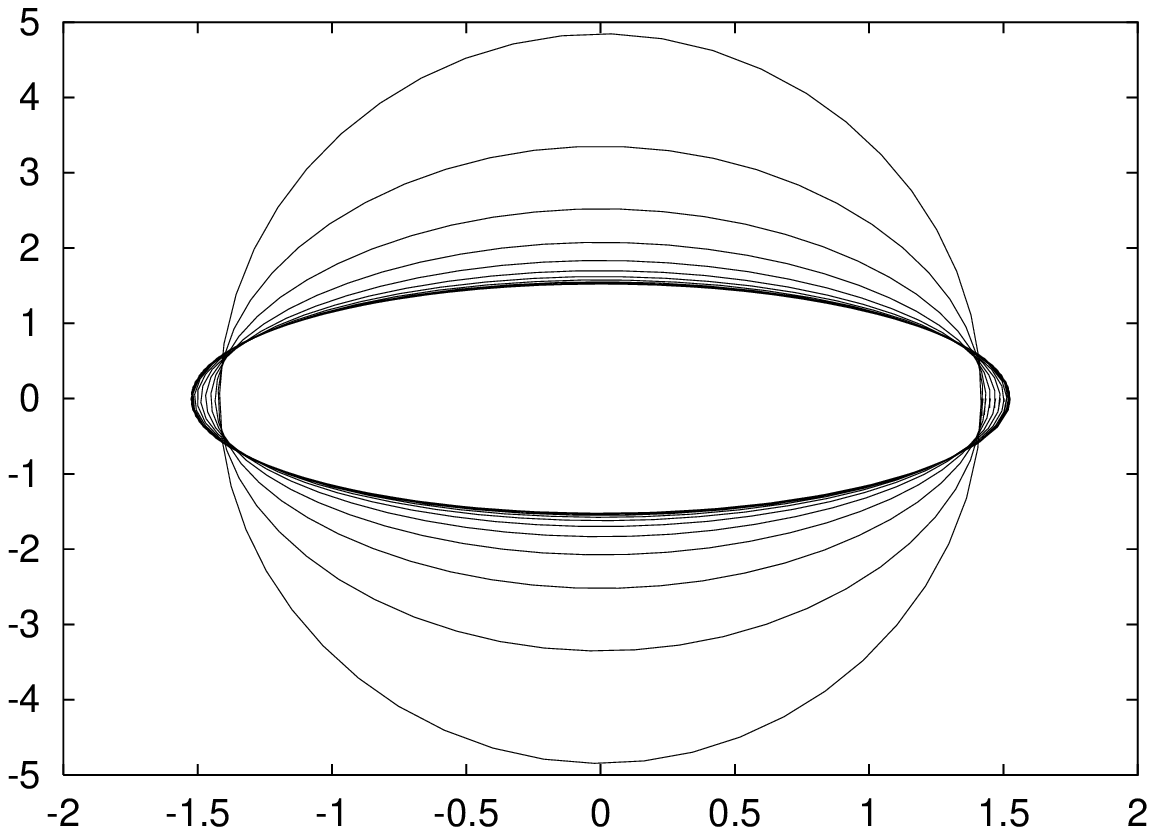,width=4.1cm,height=4.5cm}
\caption{Periodic orbits: (left) $a \times p_a$;
 (right) $\phi \times p_\phi$.}
\label{afi}
\end{center}
\end{figure}

\begin{figure}[t]
\begin{center}
\epsfig{file=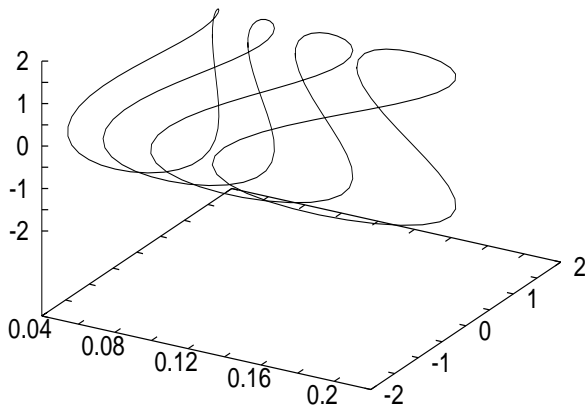,width=\columnwidth,height=6cm}
\end{center}
\caption{A three-dimensional plot ($a \times \phi \times p_\phi$) of the
periodic orbits bifurcated at $E-1.16335949$ showing the 2:1 relation between
the $a$ and $\phi$ modes.}
\label{3Dop}
\end{figure}

\begin{figure}[t]
\begin{center}
\epsfig{file=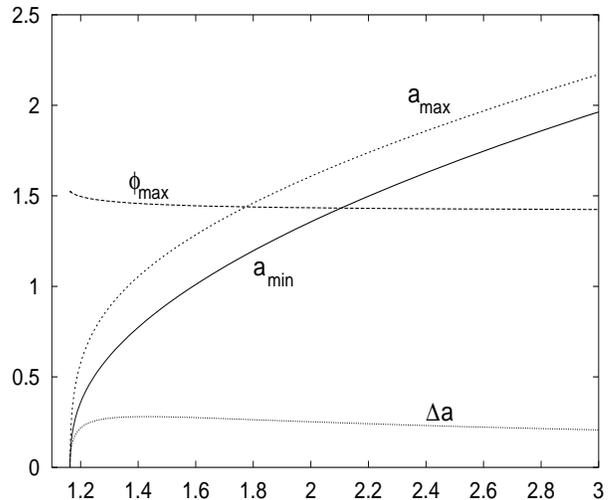,width=\columnwidth,height=7cm}
\end{center}
\caption{Range of oscillation of the scale factor $a$ and $\phi$ (symmetric
 oscillation, $\phi_{min}=-\phi_{max}$, not shown) for increasing energy. The
line $\Delta a\equiv a_{max}- a_{min}$ goes asymptotically to zero.}
\label{enamp}
\end{figure}

	We have characterized this family of noncollapsing periodic orbits
(Fig.~\ref{afi}). It presents three main features: (i) Right after the
bifurcation, the solutions present a small value for the maximum scale factor
($a_{max}\ll 1$), but a much higher $\phi$.  (ii) A $2$:$1$ resonance between
the $a$ and $\phi$  modes (Fig.~\ref{3Dop}); (iii) As the energy increases, the
scale factor  oscillates around higher mean values with decreasing amplitude
(Fig.~\ref{enamp}), while its conjugated momentum reaches a finite
maximum amplitude. The amplitude of the $\phi$ field, on the other hand, is just
slighlty reduced, while its momentum is largely amplified.  The KAM tori around
this family of periodic orbits undergoes various bifurcations as the energy
increases.

\begin{figure}[t]
\begin{center}
\epsfig{file=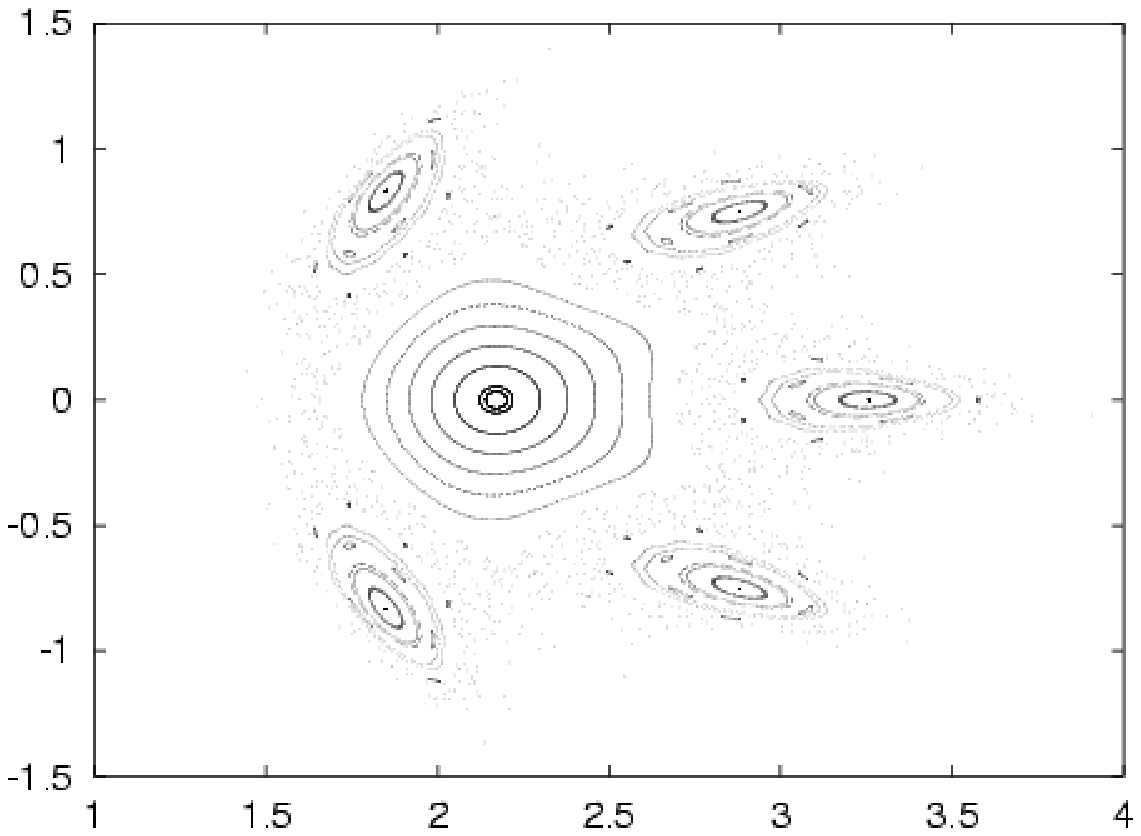,width=\columnwidth,height=8cm}
\epsfig{file=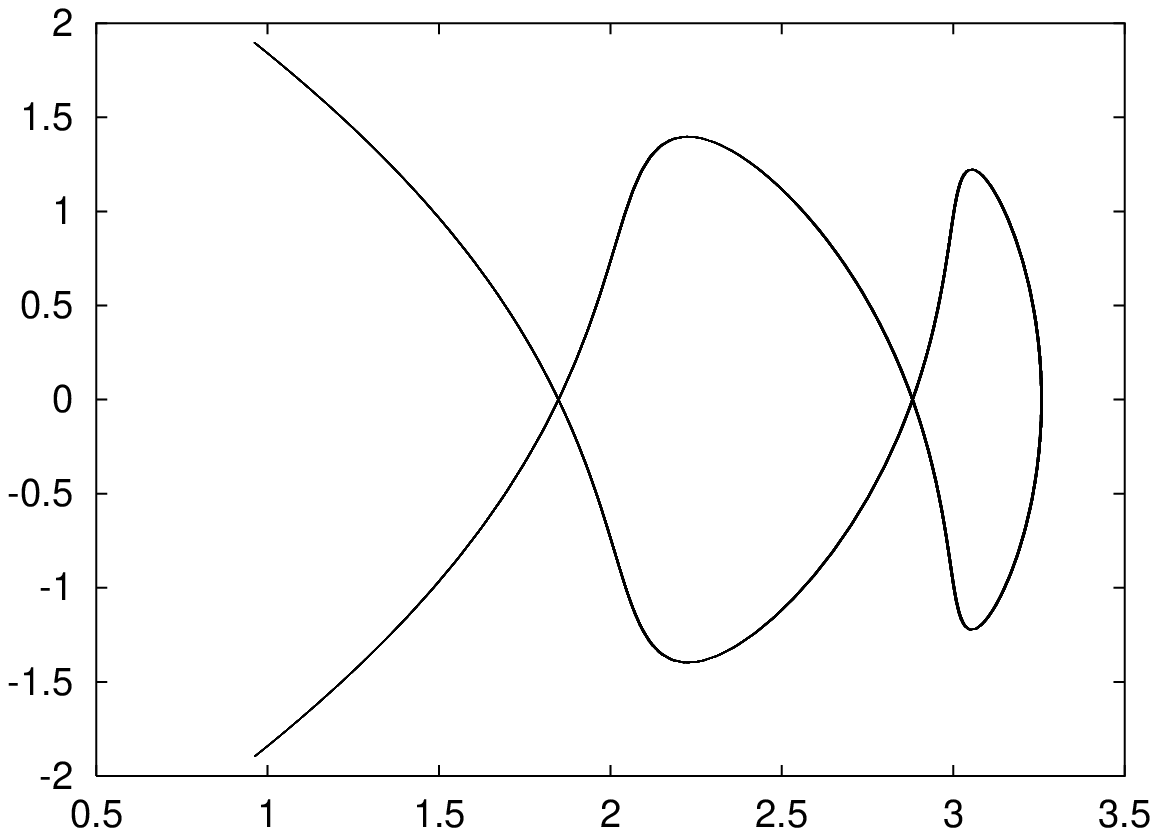,width=\columnwidth,height=8cm}
\end{center}
\caption{The bifurcation at $E=5.0$.}
\label{bif5}
\end{figure}

\begin{figure}[t]
\begin{center}
\epsfig{file=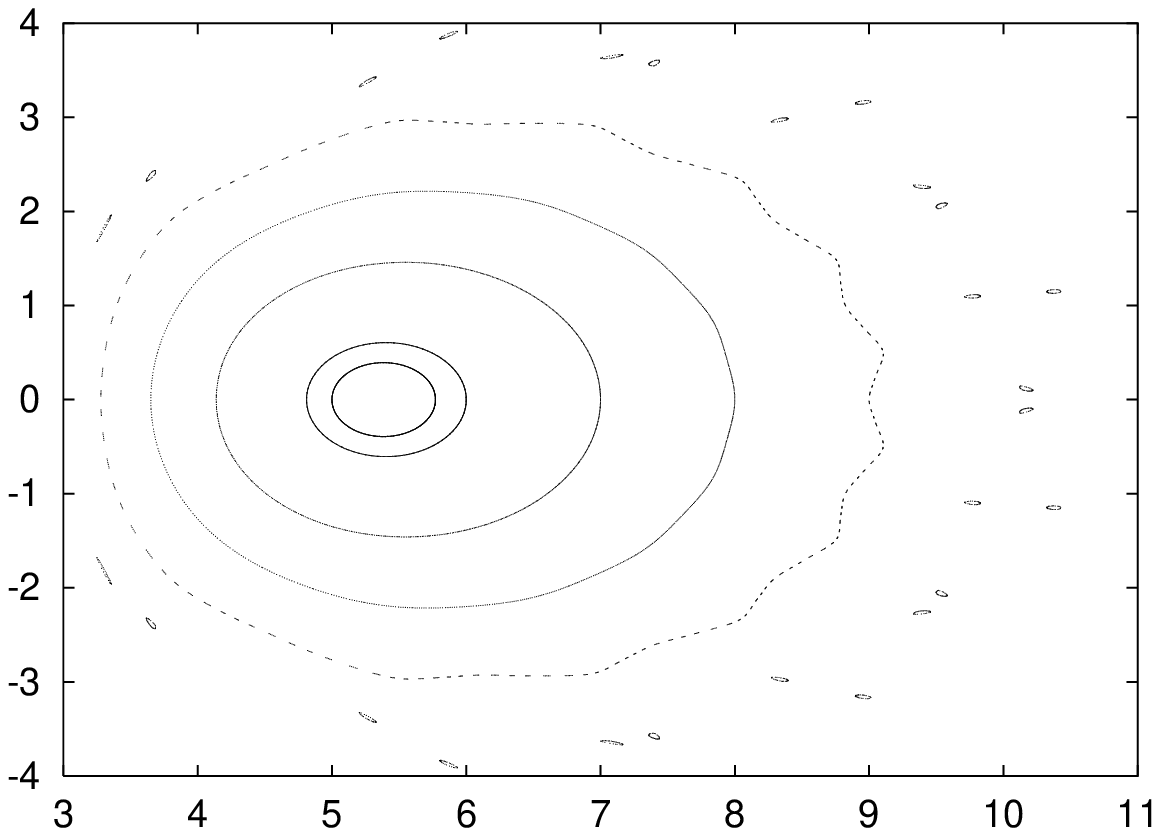,width=\columnwidth,height=7cm}
\epsfig{file=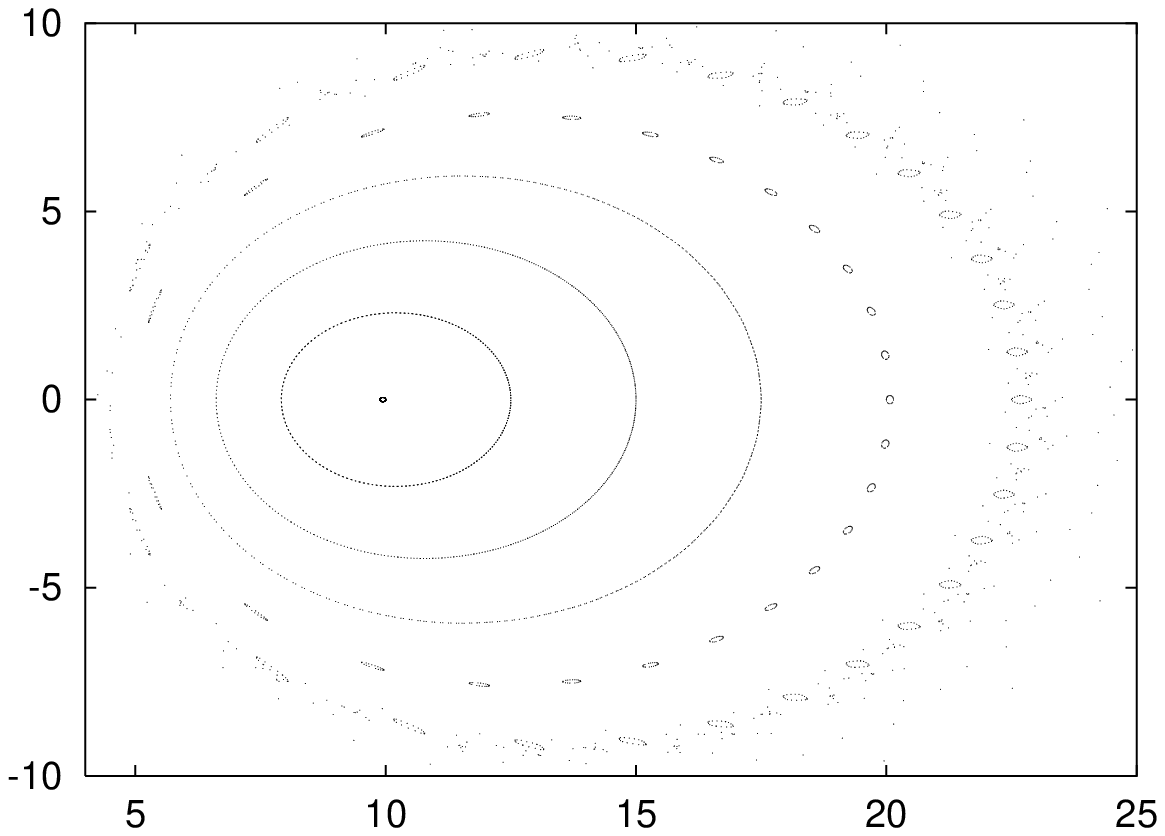,width=\columnwidth,height=7cm}
\epsfig{file=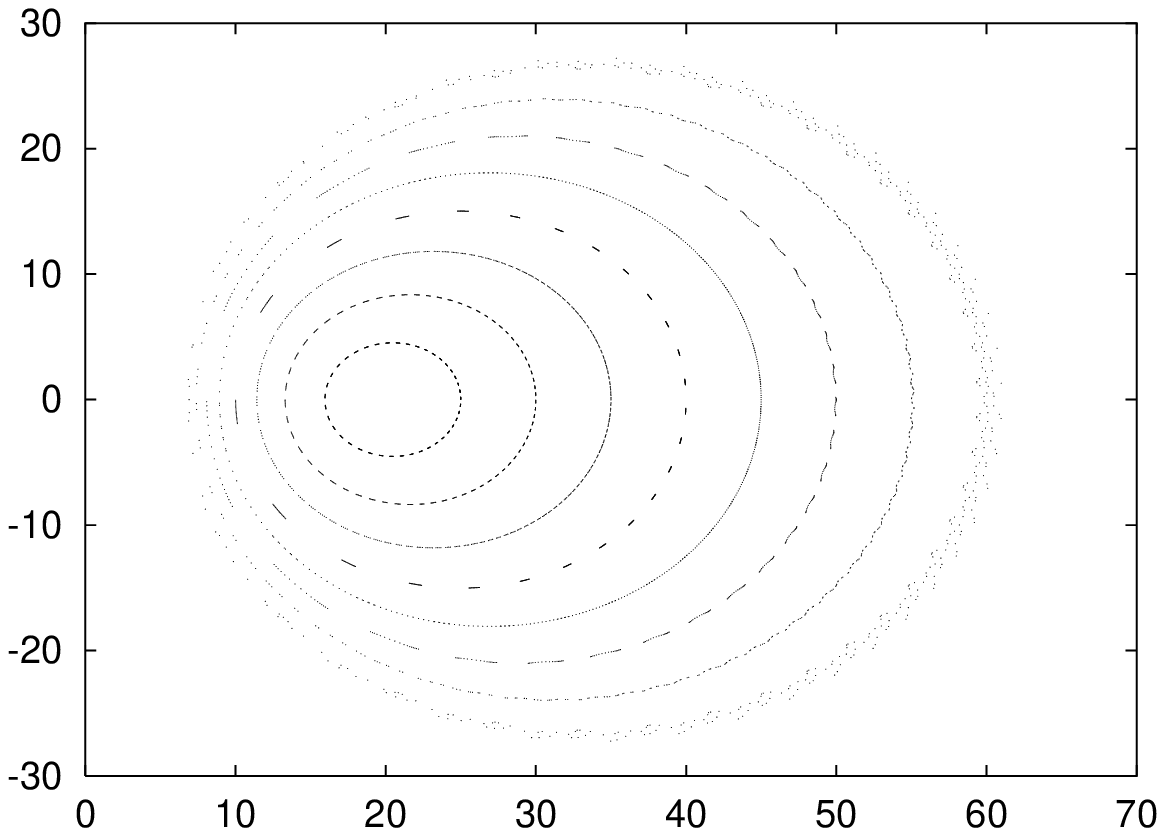,width=\columnwidth,height=7cm}
\end{center}
\caption{The regularity of KAM tori at $E=15.0$, $50.0$, and
$200$. We conjecture that there is a bifurcation of higher order at
the border of the non-colapsing structure.}
\label{50}
\end{figure}

One of these bifurcations is shown in Fig.~\ref{bif5}(a), $E=5.0$, where we can
see five islands  generating a very large stochastic zone. Fig.~\ref{bif5}(b)
exhibits a very thin noncollapsing tori of period 5. We have not systematically
investigated the width of the region of noncollapsing trajectories, but it
seems that as the energy increases the region becomes more stable and grows
proportionally to the energy.  This can be seen in Fig.~\ref{50}(a),
\ref{50}(b) and \ref{50}(c) for $E=15.0$, $E=50.0$ and $E=200.0$, respectively.
We conjecture that the islands become of increasingly higher order and are
located at the border of the noncollapsing region. As shown in Fig.~\ref{bifr},
the universe sticks to the islands for a long period of time before collapsing
-- in this particular case, it took $900<i_c<950$ Poincar\'e iterates. The
effects of chaos can indeed be observed before the collapse.

\begin{figure}[t]
\begin{center}
\epsfig{file=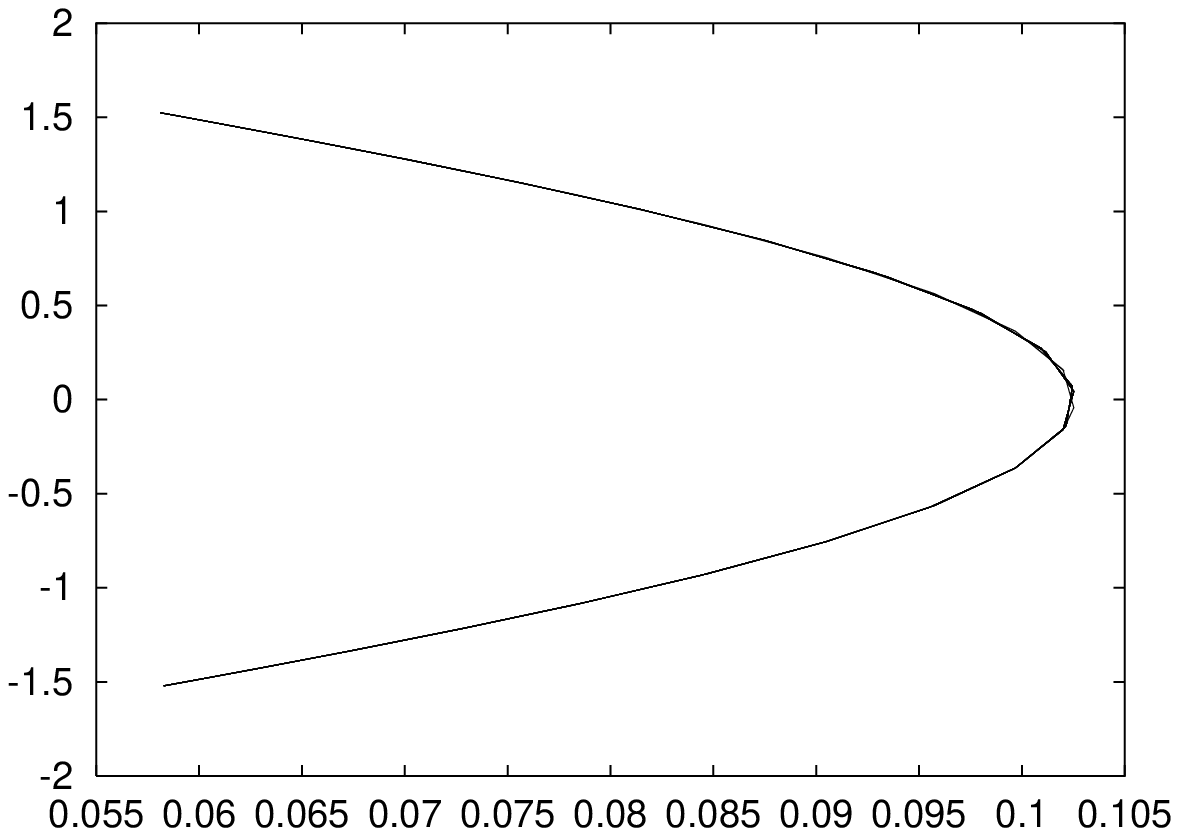,width=4.1cm,height=4cm}
\epsfig{file=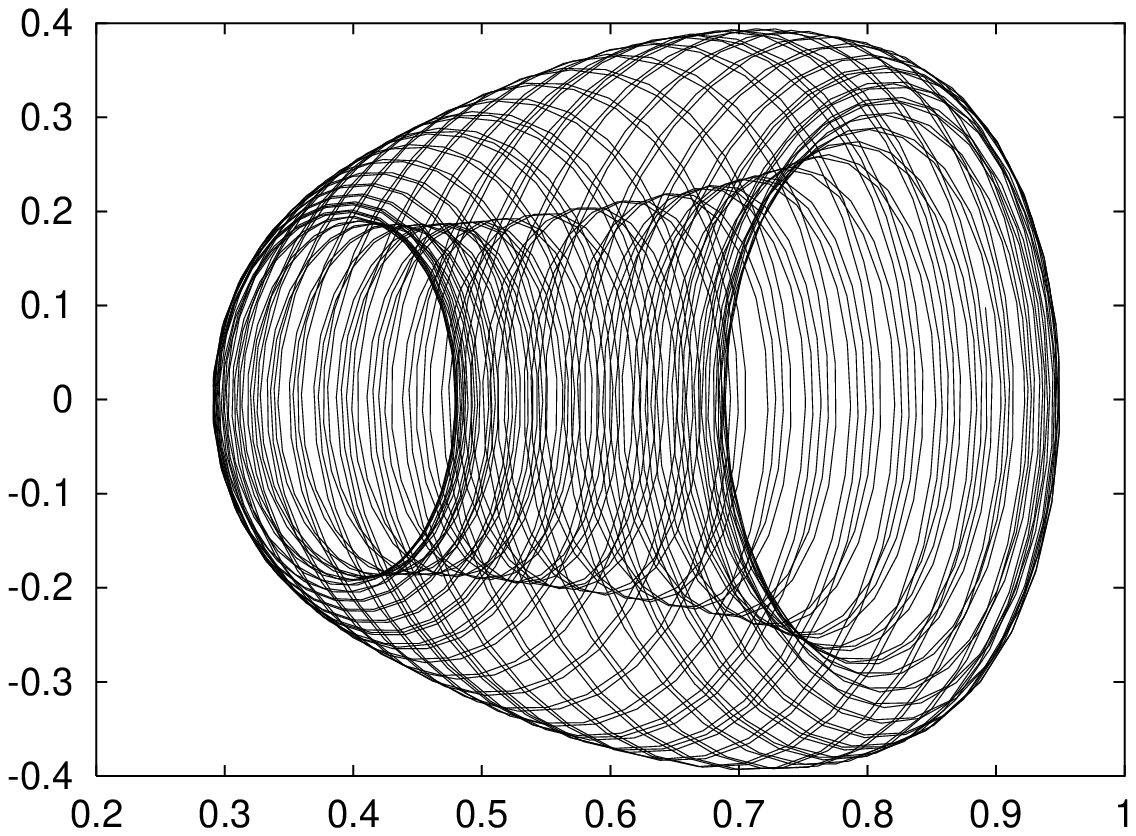,width=4.1cm,height=4cm}
\end{center}
\caption{Two noncollapsing tori for $E=1.164$ and $E=1.125$.}
\label{rpo}
\end{figure}
%

\section{Conclusions}\label{conclusion}

	The complexification of the phase space allowed us to probe the
(otherwise) complex fixed points. Their saddle-center structure is
  actually responsible for the chaotic behavior previously
described in Ref.~\cite{calzetta}. Due to the existence of heteroclinic
connections in the system, one cannot say if a given trajectory will collapse
or escape to infinity.

       The most important outcome of our procedure is to discover that, as the
origin  is reached by the heteroclinic intercepting loops of the stable and
unstable manifolds in the analytical continuation, a bifurcation is
created  at that point, both in the extension and in the real counterpart.
We have also shown that, as a consequence,  there are KAM structures formed by
trajectories which will never collapse after $E=1.16335949$, and they
are persistent to very high values of the energy.

\section{Acknowledgment}

The authors wish to thank Mario B. Matos for interesting discussions. S.E.J.
thanks the UFRJ for the kind and warm hospitality, where most of this work was
accomplished. We are also indebted to Funda\c c\~ao Jos\'e Bonif\'acio (FUJB)
for partial computational support. S.E.J. was partially supported by CNPq.


\end{document}